\documentclass[paper]{JHEP3}
\pdfoutput=1
\usepackage{amsmath,amssymb,amsthm,amscd,graphicx}
\usepackage{wrapfig}

\theoremstyle{definition}

\newcommand{\tr}{{\rm Tr}}
\newcommand{\re}{{\rm e}}

\renewcommand{\d}{\partial}

\newcommand{\be}{\begin{equation}}
\newcommand{\ee}{\end{equation}}
\newcommand{\ba}{\begin{aligned}}
\newcommand{\ea}{\end{aligned}}
\newcommand{\ben}{\begin{eqnarray}\displaystyle}
\newcommand{\een}{\end{eqnarray}}
\newcommand{\nn}{\nonumber}

\newcommand{\sectiono}[1]{\section{#1}\setcounter{equation}{0}}

\newdimen\tableauside\tableauside=2.0ex
\newdimen\tableaurule\tableaurule=0.8pt
\newdimen\tableaustep
\def\phantomhrule#1{\hbox{\vbox to0pt{\hrule height\tableaurule width#1\vss}}}
\def\phantomvrule#1{\vbox{\hbox to0pt{\vrule width\tableaurule height#1\hss}}}
\def\sqr{\vbox{%
  \phantomhrule\tableaustep
  \hbox{\phantomvrule\tableaustep\kern\tableaustep\phantomvrule\tableaustep}%
  \hbox{\vbox{\phantomhrule\tableauside}\kern-\tableaurule}}}
\def\squares#1{\hbox{\count0=#1\noindent\loop\sqr
  \advance\count0 by-1 \ifnum\count0>0\repeat}}
\def\tableau#1{\vcenter{\offinterlineskip
  \tableaustep=\tableauside\advance\tableaustep by-\tableaurule
  \kern\normallineskip\hbox
    {\kern\normallineskip\vbox
      {\gettableau#1 0 }%
     \kern\normallineskip\kern\tableaurule}%
  \kern\normallineskip\kern\tableaurule}}
\def\gettableau#1{\ifnum#1=0\let\next=\null\else
\squares{#1}\let\next=\gettableau\fi\next}

\title{Mirror of the refined topological vertex from a matrix model}

\author{Bertrand Eynard$^{\S,\dagger}$, Can Koz\c{c}az${}^\dagger$
\\
$^{\S}$
Institut de Physique Th\'eorique, IPHT/DSM/CEA Saclay\\
F-91191 Gif-sur-Yvette Cedex, France
\\\\ 
$^\dagger$
PH-TH Division, CERN\\
Gen\`eve, CH-1211 Switzerland 
}

\preprint{CERN-2011/182 \\
IPHT-T11/xxx}
\abstract{We find an explicit matrix model computing the refined topological vertex, starting from its representation in terms of plane partitions. We then find the spectral curve of that matrix model, and thus the mirror symmetry of the refined vertex. With the same method we also find a matrix model for the strip geometry, and we find its mirror curve. The fact that there is a matrix model shows that the refined topological string amplitudes also satisfy the remodeling the B--model construction.}    

\begin{document}
\tableofcontents
\section{Introduction}
\par{The topological string theory has provided a lot of excitement in both physics and mathematics. It comes in two different versions, A- and B-models, based on the different twists of the worldsheet theory. Many techniques have been developed to compute their amplitudes. The techniques manifest different features of the topological string theory and are usually complementary to each other. For the present work, on the A-model side, the \textit{refined topological vertex}  \cite{Iqbal:2007ii} is the method of interest, and on the B-model side the \textit{remodeling approach} \cite{Bouchard:2007ys} based on the topological recursion. We want to connect these two approaches to shed some light in the refinement of the mirror symmetry.}
\medskip

\par{The topological vertex formulation gives the all genus amplitude for the toric Calabi-Yau threefolds. This includes geometries which engineer the ${\cal N}=2$ $SU(N)$ gauge theories in 4$d$ in type IIA compactifications \cite{Katz:1996fh}. The genus zero amplitude of the topological string theory gives the prepotential of the gauge theory. Higher genus amplitudes compute the gravitational coupling to the gauge theory. The genus expansion of the topological string theory has been connected to the instanton calculus on the gauge theory side and allowed what is now called the \textit{refinement} of the topological string theory. The refined topological vertex has been motivated to understand the A-model side of the extension.  }
\medskip

\par{On the other hand, the connection between the topological string theory and matrix models has been under extensive exploration for a long time starting with \cite{Dijkgraaf:2002fc,Dijkgraaf:2002vw}. Both open and closed topological string amplitudes have been computed for various backgrounds using matrix models for both the topological A and B models. The integrability of topological strings on these backgrounds is one of the immediate consequences of this correspondence. More recently, the topological recursion is developed to solve the loop equations of a matrix model in a recursive manner \cite{eynardFg1MM}. The procedure starts with a spectral curve and basic ingredients associated with it. The spectral curve is obtained by the large $N$ limit of the Schwinger-Dyson equations of the matrix model, then once the spectral curve is known, the topological recursion generates a set of meromorphic forms and complex numbers only based on the geometric properties of the spectral curve. Later the formalism is extended beyond the matrix model by choosing algebraic curves that are not coming from the large $N$ limit \cite{EO}. Many remarkable properties are shown.    }
\medskip

\par{The explicit relationship between the topological string theory on toric geometries and the topological recursion has appeared for the first time in \cite{Marino:2006hs}. It has been based on the observation that both the local B-model topological string theory amplitudes and the general solutions of matrix models can be related to chiral bosons living on a Riemann surface. The Riemann surface in question is the reduction of the Calabi-Yau threefold or the spectral curve, respectively. The mirror geometry of toric threefolds can be reduced to a Riemann surface as well. The meromorphic differentials are constructed on this mirror curve. This initial attempt is completed in \cite{Bouchard:2007ys}. This framework allows one to compute open and closed B-model amplitudes unambiguously at any point in the moduli space, but still perturbative in the string coupling. The integrals of the differentials correspond to the framed open amplitudes and the invariants are related to the closed amplitudes of the A-model, after taken the mirror map into account. However, this approach is restricted to the unrefined case.}
\medskip

\par{Our aim is to extend the remodeling to match the refined A-model computations. More precisely, we are after the refined spectral curve, \textit{i.e.}, the \textit{refined mirror curve}, to perform  the topological recursion to give the refined topological string amplitudes. We will first construct a matrix model for the refined topological vertex and the refined topological strip. Then we will find the spectral curves of these matrix models. The first step is straightforward considering that the refined topological vertex has been defined through plane partitions and a matrix model for plane partitions with arbitrary weights has been already constructed \cite{eynard1}. For the second step we will benefit from the results of \cite{eynchain,eynpratschain} where the spectral curve of the resulting chain matrix model has been found using saddle point approximation. We find explicit expression for the spectral curves as a function of $\beta=-\epsilon_{1}/\epsilon_{2}$.  }
\medskip

\par{The closed refined strip computation has been connected to (chiral) $A_{N-1}$ Toda theory ($T_{N}$-theory) three-point function with one of the three primary fields of a special type in \cite{Kozcaz:2010af} using the AGTW correspondence \cite{Alday:2009aq,Wyllard:2009hg}. Our construction for the refined strip implies a matrix model for the three-point function, hence, manifestly integrable structure for them. In \cite{Benini:2009gi} a analogous proposal for three-point functions of a 5$d$ version Toda theory has been made based on the five-brane webs. The brane configurations are related to the toric spaces \cite{Leung:1997tw}. For $T_{2}$, the corresponding geometry is  the so-called closed vertex geometry and the chiral three-point function has been shown to agree with the 4$d$ limit of the refined topological vertex result. Higher rank theories involve more complicated geometries and it is not obvious how to take the 4$d$ limit. Our matrix model construction can be extended similar to \cite{Eynard:2010dh} to capture all possible toric geometries. It will allow to connect matrix models to the proposal in \cite{Benini:2009gi}. }
\medskip
\par{The paper is organized as fallows: In section 2 we review the basic definition of the plane partitions and the description of plane partition in terms of non-intersecting paths. This description is very crucial to write down the corresponding matrix model. In section 3 we overview the topological vertex formulation. In section 4 we remind the basic ingredients for the corresponding matrix model. We also make the connection to the refined topological vertex. In section 5 we extend our matrix model to cover the strip geometry in the refined case. In section 6 we focus on finding the spectral curve of the matrix models and the corresponding limiting shapes. In section 7 we comment on the use of topological recursion. In section 8 we work out some examples and certain limit of them. In section 9 we finish with the conclusions. In the appendix we collect some useful formulae used in the derivations.    }

\sectiono{Plane partitions }

\par{In this section we want to review the basics about the 2$d$, also known as Young diagrams and 3$d$ partitions, the so-called plane partitions. This section should serve more as fixing the notation rather than a detailed introduction. At the end of this section we want to remind how the plane partitions can be represented in terms of non-intersecting paths. In other words, they can be visualized as particles moving from one end to the other in discrete times and are only allowed to make jumps by $\pm1/2$ to the next time slice such that no two particles can occupy the same position. }

\subsection{Partitions}

\par
\begin{figure}[h]
\begin{center}
\includegraphics[scale=1.]{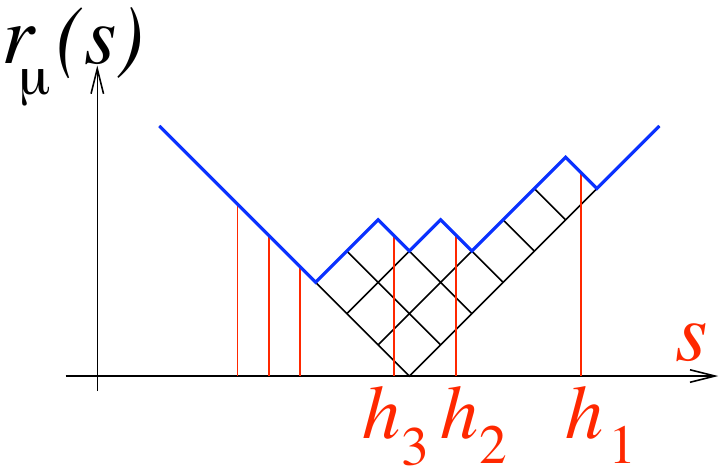}
\end{center}
\caption{The 2$d$ partition rotated by 45 degrees, the horizontal coordinate of the descending edge is labelled by $h_{k}$. The upper blue line is the profile of the partition.\label{figpartrotated}}
\end{figure}

{Before defining the plane partition it is useful to review the 2$d$ partition or the so-called Young diagrams. A sequence of positive integers of decreasing order $\mu=\{\mu_{1}\geq\mu_{2}\geq\mu_{3}\geq\mathellipsis \}$} defines a 2$d$ partition. By placing $\mu_{i}$ boxes on the $i^{th}$ position the 2$d$ partition can be pictorially visualized. 

$$\mu=\{6\geq 3\geq 2\} \qquad \longrightarrow \quad\tableau{2 3 6}\quad .$$
It is also interesting to represent a 2$d$ partition $\mu$, rotated by 45 degrees, and define:
\be
h_k = \mu_k-k+a,
\ee
where $a$ is a fixed reference point, which we shall choose later on.
$h_k$ is the horizontal coordinate of the end of row $\mu_k$ rotated by 45 degrees, see figure \ref{figpartrotated}, \textit{i.e.}, the $h_k$'s correspond to the descending edges of the rotated partition.
We have
\be
h_1>h_2>h_3>\dots > h_N\geq a-N.
\ee
The profile of a partition $\mu$ is the function $r_\mu(s):\mathbb R\to \mathbb R$, defined by
\be\label{defprofile}
r_\mu(s) = \frac 1 2 |s| + \frac 1 4 \,\sum_{i} \,|s-(\mu_i-i+1)|+|s+i|-|s-(\mu_i-i)|-|s+i-1|, 
\ee
which is a piecewise affine function with slopes $r'_\mu(s)=\pm \frac 1 2$, see figure  \ref{figpartrotated}.
The slopes at integer $s$ might be ill defined, but they are always well defined at half-integers.
We define $S_-(\mu)$ to be the set of descending edges, \textit{i.e.}, the set of $h_k$'s:
\be
S_-(\mu) = \left\{ s\in\mathbb Z\, | \, r'_\mu(s-\frac 1 2)=-\frac 1 2\right\} = \Big\{h_1,h_2,h_3,\dots\Big\}
\ee
and we also define the set of ascending edges:
\be
S_+(\mu) = \left\{ s\in\mathbb Z\, | \, r'_\mu(s-\frac 1 2)=+\frac 1 2\right\} =\mathbb Z\setminus S_-(\mu).
\ee

\subsection{Plane partitions}

\par{A plane partition is the 3$d$ generalization of Young diagrams. It is defined as an array of non-negative integers satisfying}
\be\label{defpplane}
\pi_{i,j}\geq \pi_{i+r,j+s},\qquad r,s\geq 0.
\ee
\begin{figure}[h]
\begin{center}
$$\includegraphics[scale=.3]{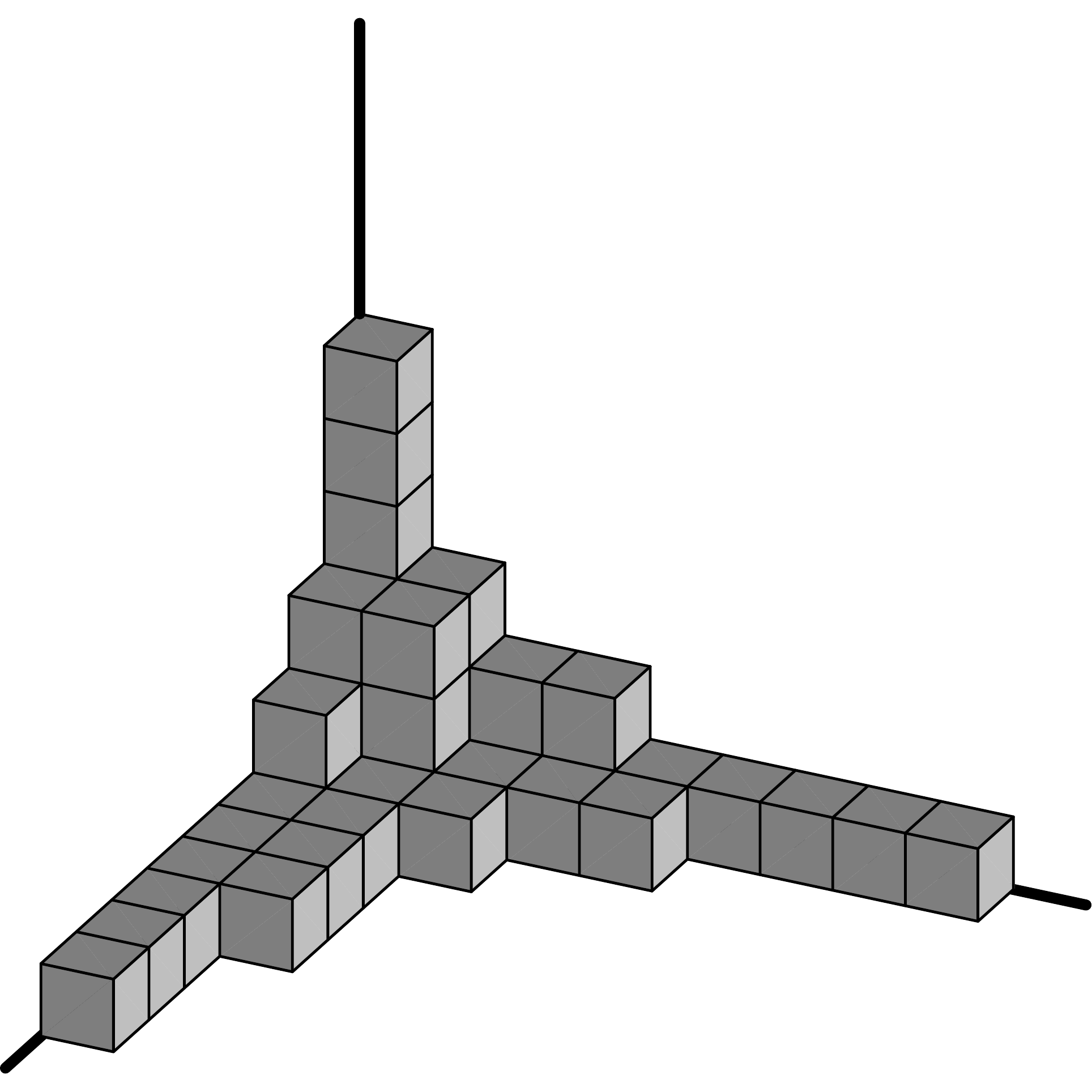}$$
$$
 \includegraphics[scale=0.55]{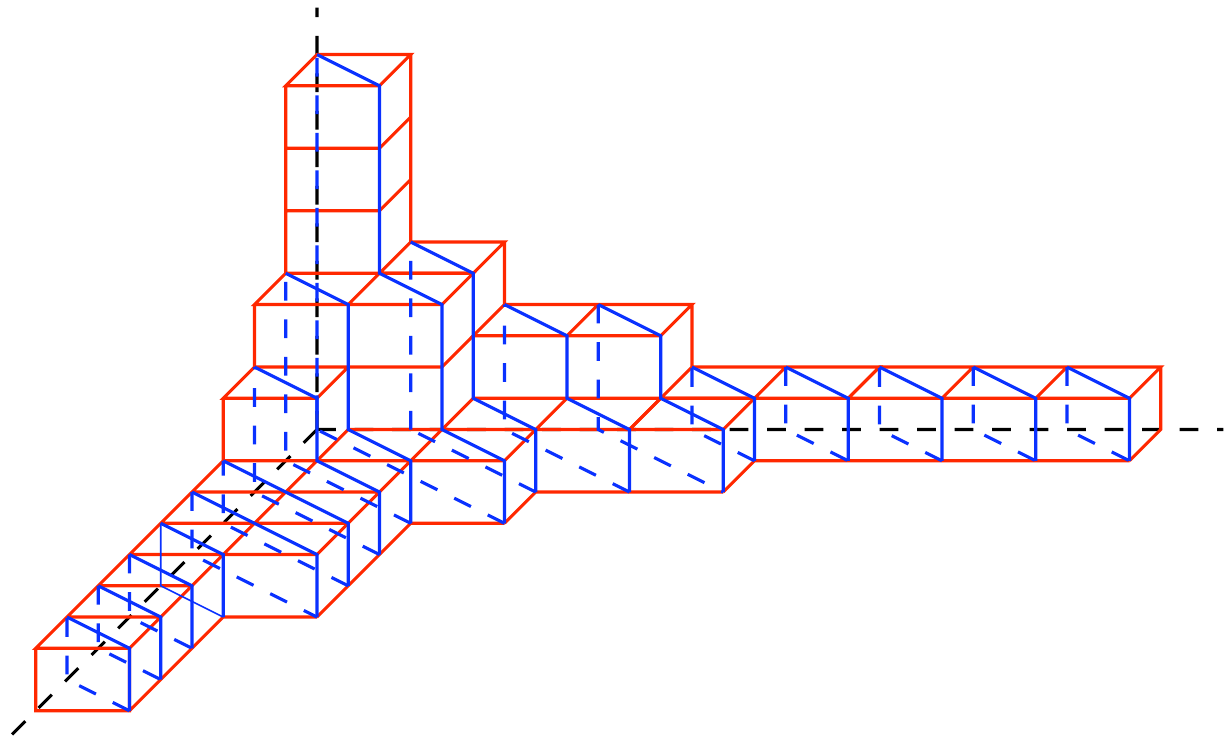}
 \qquad
 \includegraphics[scale=0.55]{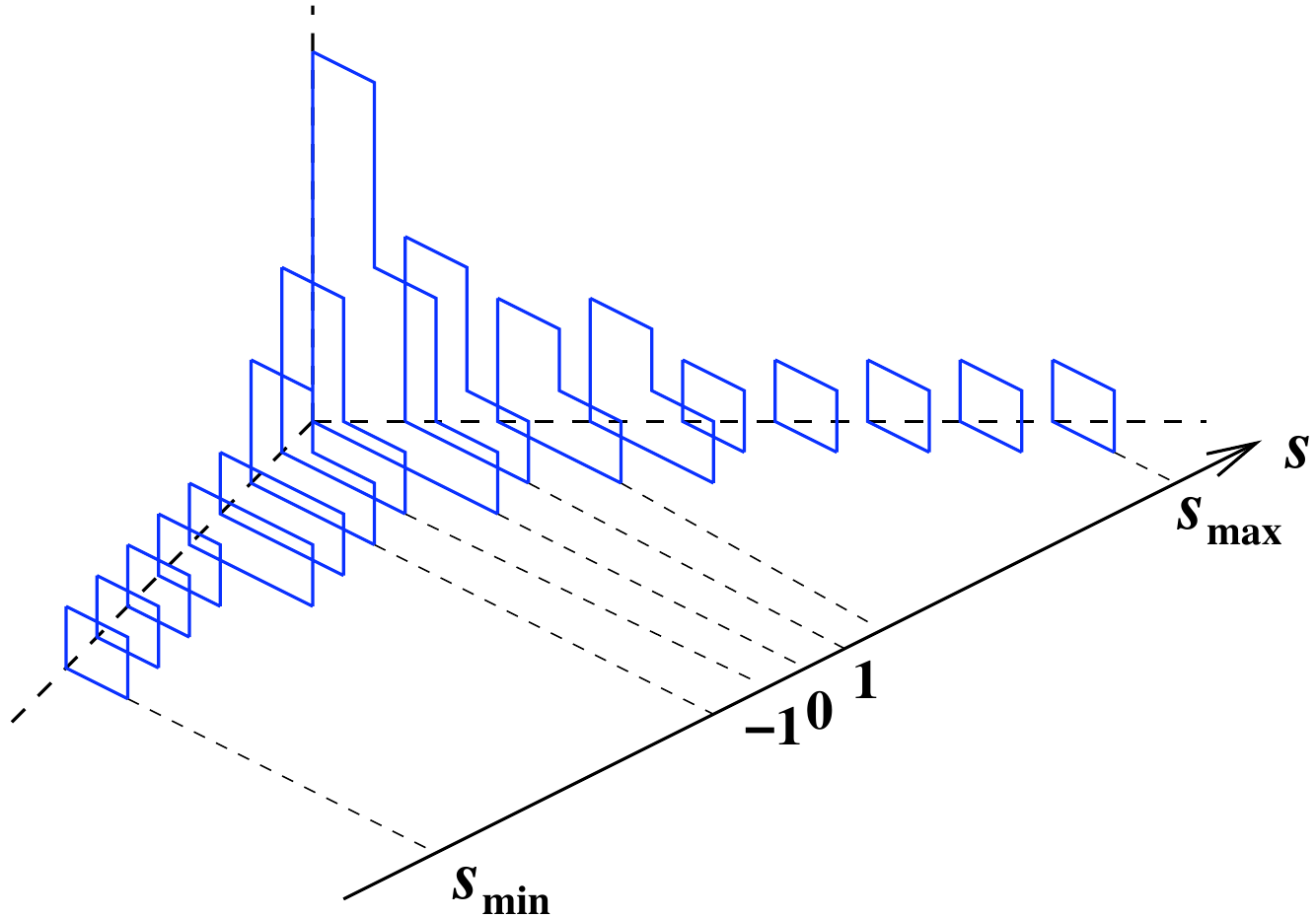}
$$
\end{center}
\caption{An example for a plane partition, and its slices.\label{fig3dslices}}
\end{figure}
\par{The plane partition can be pictorially represented by putting $\pi_{i,j}$ boxes over each square labeled by $(i,j)$. It can also be viewed as a crystal growing in the corner of a room: new boxes can attach only in corners. 
A plane partition $\pi$ can also be represented by a set of 2$d$ partitions $\eta(s)=\{\eta_1(s)\geq \eta_2(s)\geq \dots \}$, labelled by $s\in \mathbb Z$. $\eta(s)$ is called the $s^{\rm th}$ diagonal slice (see figure \ref{fig3dslices}), defined by}
\be\label{defslices}
\pi = \{\dots,\eta(-1),\eta(0),\eta(1),\eta(2),\dots\},
\qquad {\rm where}\,\,\, \eta(s)=\,
\left\{\begin{array}{ll}
\eta_k(s) = \#\{i\,|\,\,\pi_{i,s+i}\geq k\} & \quad {\rm if}\,\, s\leq 0 \cr
\eta_k(s) = \#\{j\,|\,\,\pi_{s+j,j}\geq k\} & \quad {\rm if}\,\, s\geq 0 
\end{array}\right. .
\ee

\subsection{Plane partitions and non--intersecting paths}

\par{Plane partitions are known to be expressed as a model of self-avoiding jumping particles. 
Consider a bounded plane partition, with three boundaries given by three partitions $\lambda,\, \mu$ and $\nu$, located  at respective distances $N_\lambda=N$, $N_\mu=-s_{\rm min}$, $N_\nu=s_{\rm max}$ from the origin, see figure \ref{self}. 

The plane partition can also be represented as the set of its diagonal slices $\eta(s)$, $s_{\rm min}\leq s\leq s_{\rm max}$, as defined in \eqref{defslices}. Since $\pi$ has an height of $N$, each slice $\eta(s)$ is a $2d$ partition with at most $N$ non--empty rows.
\medskip

As we have seen above, a 2$d$ partition $\eta(s)=\{\eta_i(s)\}$, with rows $\eta_1(s)\geq \eta_2(s)\geq\dots\geq \eta_N(s)\geq 0$, can be encoded by its 45 degrees rotated version, \textit{i.e.}, the set $\{h_i(s)\}_{i=1,\dots,N}$ with
\be\label{defhis}
h_i(s)=\eta_i(s)-i+N+\frac 1 2 \,|s|,
\ee
where we have chosen the origin (\textit{i.e.} $\eta_N(s)=0$) to be on the line $h=\frac 1 2\,|s|$.
$h_i(s)$ is the position of the $i^{\rm th}$ vertical edge of $\eta(s)$, along the $s^{\rm th}$ vertical line.
\medskip

The fact that $\pi$ has boundary $\lambda$ from above implies that 
\be
h_1(s)> h_2(s)>\dots> h_N(s)\geq r_\lambda(s),
\ee
where $r_\lambda(s)$ is the profile of the partition $\lambda$ bounding $\pi$ from the top, defined in \eqref{defprofile}. We obtain the 2$d$ partitions by slicing a plane partition, this imposes the following condition among adjacent slices
\be\label{cond}
h_i(s+1)=h_i(s)\pm \frac 1 2.
\ee

It is customary to represent the partition $\eta(s)$ by a set of $N$ ``particles" living on a 1$d$ line (coordinate $h$), the position of the $i^{\rm th}$ particle at time $s$ is $h_i(s)$. Those particles follow a time evolution (the time is $s$, going from the left side to the right side of $\pi$), such that at all times $s$ the particles can never occupy the same position and can never cross (indeed $h_1(s)>h_2(s)>\dots>h_N(s)$), and move by jumps of $\pm \frac 1 2 $. See figure figure \ref{part}.

\begin{figure}[h]
\begin{center}
\includegraphics[scale=0.7]{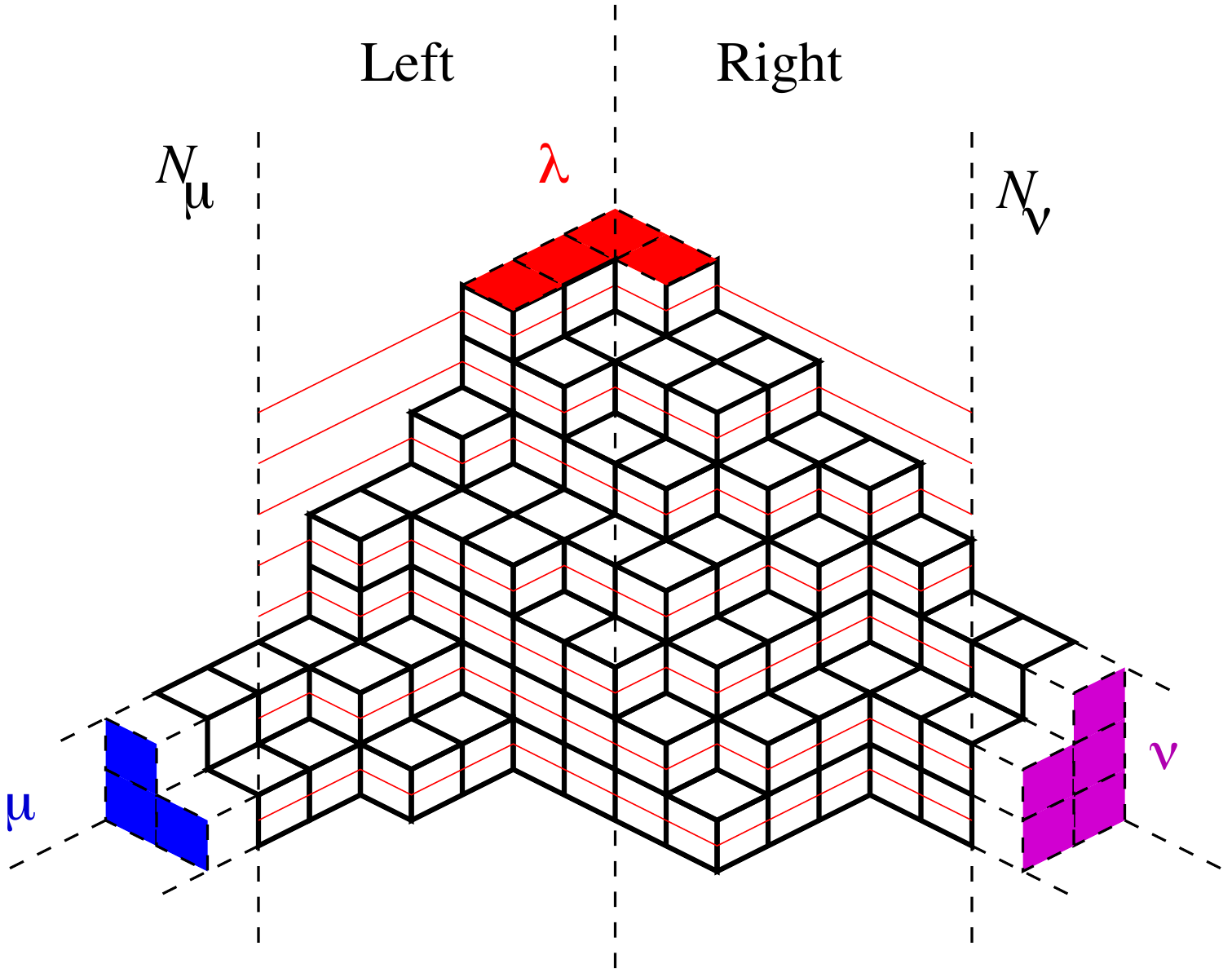}
\end{center}
\caption{A plane partition with boundaries labeled by $\lambda,\, \mu$ and $\nu$. The red lines represent the condition (\ref{cond}).}\label{self}
\end{figure}

The trajectories of those $N$ particles are $N$ non-intersecting
 lines connecting the left boundary to the right boundary. The intersection of the lines with the (vertical) time lines are the location $h_{i}(s)$ of the particles at each integer time $s=s_{\rm min},\mathellipsis, s_{\rm max}$. Each of the $N$ lines will represent the world-line of a particle discretely moving in one dimension, see figure \ref{part}.  
\medskip

\begin{figure}[h]
\begin{center}
\includegraphics[scale=0.6]{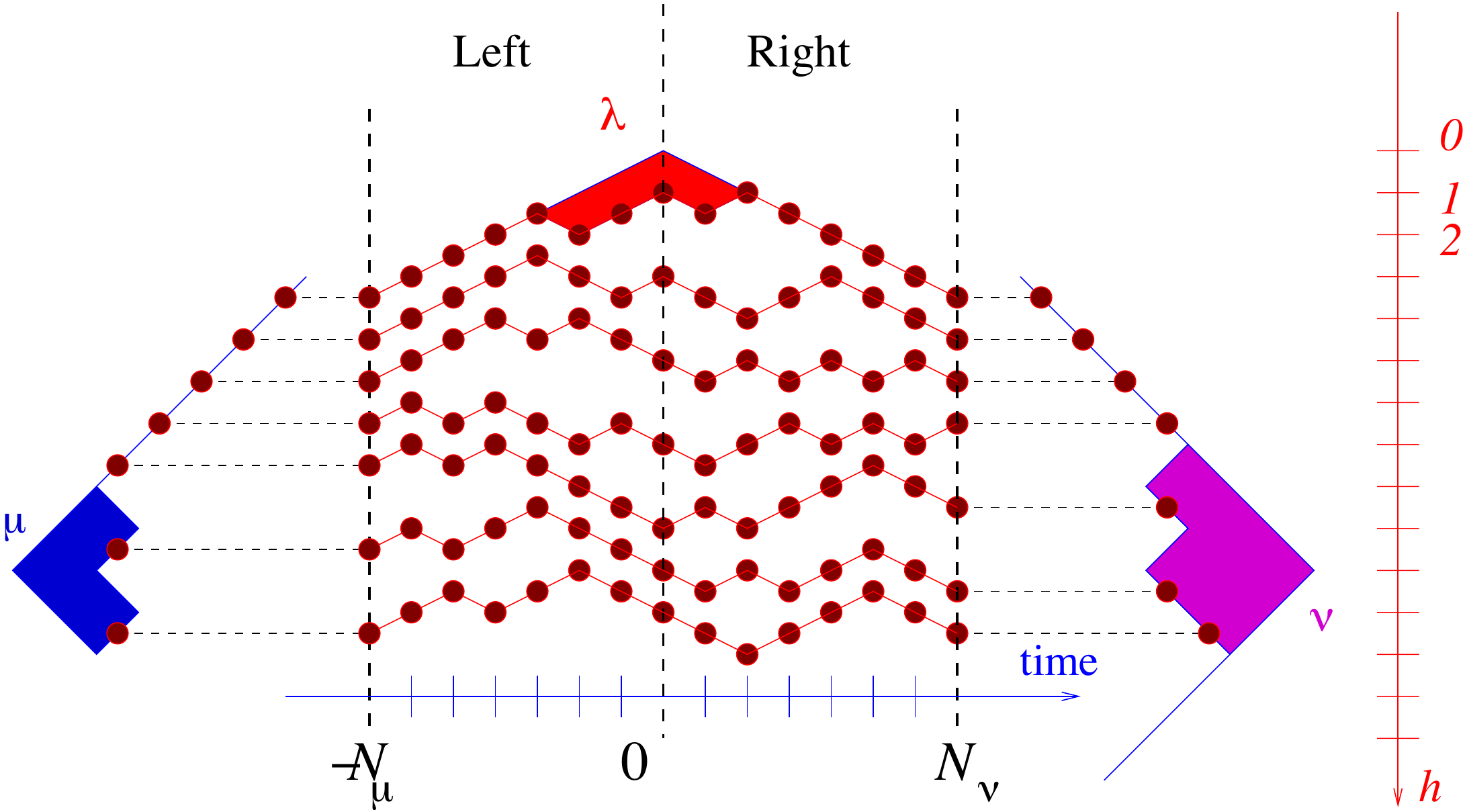}
\end{center}
\caption{For each plane partition there is a set of trajectories of particles moving in 1$d$ from the left boundary to the right one, subject to the boundary conditions. \label{part}}
\end{figure}

A plane partition configuration, is thus encoded by the data of the positions $h_i(s)$ of $N$ particles ($i=1,\dots, N$) at times $s$ with $s_{\rm min}\leq s\leq s_{\rm max}$, whose initial positions $h_i(s_{\rm min})$ and final positions $h_i(s_{\rm max})$ are fixed:
\be
\pi\equiv\{h_i(s)\}_{i=1,\dots,N\, \,\,\, , \, s_{\rm min}\leq s\leq s_{\rm max}}
\qquad {\rm such\, that}
\qquad
\left\{\begin{array}{l}
h_1(s)>h_2(s)>\dots > h_N(s)\geq r_\lambda(s) , \cr
h_i(s+1) = h_i(s) \pm \frac 1 2 , \cr
h_i(s_{\rm min}) = \mu_i-i+N+\frac {|s_{\rm min}|}{2}, \cr
h_i(s_{\rm max}) = \nu_i-i+N+\frac {|s_{\rm max}|}{2}. 
\end{array}\right.
\ee

\setcounter{equation}{0}
\section{The topological vertex formalism}
\par{The topological string theory offers a simplified framework to understand the physical string theory. It shares many important phenomena with the string theory but in a computational more tractable manner. The topological string amplitudes appear in the 4$d$ effective action in string theory compactifications. They are also generating functions for BPS states in various supersymmetric theories. }\medskip

\par{Supersymmetric gauge theories can be constructed within the framework of string theory in different ways. One of the most powerful constructions is the so-called geometric engineering \cite{Katz:1996fh}. The gauge theory data such as the gauge group, the matter content and the amount of supersymmetry is encoded in the geometry of the space $X$ that is used to compactify type IIA string theory from 10$d$ down to 4$d$ or M-theory from 11$d$ to 5$d$. The geometric engineering gives a recipe to determine this space $X$ based on the gauge theory data. One of the most important advances in string theory is the realization that the genus zero amplitude of the topological string theory on $X$ computes the prepotential of the corresponding gauge theory. Higher amplitudes compute the gravitational corrections. More precisely, in type IIA compactification the effective action in the 4$d$ transverse space includes F-terms describing the coupling of the self and anti-self dual graviphoton fields to self and anti-self dual contractions of the Riemann tensor }
\be
\int d^{4}x\, F_{g}(\omega)(R_{+}^{2}F_{+}^{2g-2}+R_{-}^{2}F_{-}^{2g-2}),
\ee
where $F_{g}(\omega)$ is the genus $g$ topological string amplitude. It depends on the K\"{a}hler moduli $T_{\beta}=\int_{\beta}\omega$ of the internal space. They are generating functions for genus $g$ Gromov-Witten invariants and are computed by performing integral over the moduli space of Riemann surfaces. Turning on only the self-dual graviphoton field we get corrections to the $R_{+}^{2}$ term 
\be
\left( \sum_{g=0}^{\infty}F_{+}^{2g-2}F_{g}(\omega) \right)R_{+}^{2},
\ee  
where $F_{+}$ plays the role of the topological string coupling constant. The partition function is defined
\be
Z(\omega,g_{s})=e^{F(\omega,g_{s})}=\exp\left( \sum_{g=0}^{\infty}g_{s}^{2g-2}F_{g}(\omega) \right).
\ee
The target space point of view offers a more physical interpretation of the topological string partition function \cite{Gopakumar:1998ii, Gopakumar:1998jq}. If we compactify type IIA on $X$, in the remaining 4d space we will have BPS particles which come from D2 branes wrapping  two cycles $\beta\in H_{2}(X,{\mathbb Z})$ in $X$. The mass of each particle is given by the area of the cycle $T_{\beta}$ the brane is wrapping. There is a caveat due to the bound D0 branes to D2 branes. Let us lift to the M theory, then a D2 brane  becomes an M2 branes wrapping the corresponding two cycle. An M2 brane comes with momenta along the circle direction. We can interpret these Kaluza-Klein modes in the M-theory with the D0 branes. The mass of the M2 brane with the momentum $n$ is shifted from $T_{\beta}$ to $T_{\beta}+2\pi i n /g_{s}$. In addition, it carries quantum numbers under the little group of massive particles in 5$d$, $(j_{L},j_{R})\in SU(2)_{L} \times SU(2)_{R}=SO(4)$. Let us denote the degeneracy of the particles of given quantum numbers by $N_{\beta}^{(j_{L},j_{R})}$. The free energy can be written as contributions coming from all such particles:
\be
F(\omega;q)=\sum_{\beta\in H_{2}(X,{\mathbb Z})}\sum_{k=1}^{\infty}\sum_{j_{L}}(-1)^{2j_{L}}N_{\beta}^{(j_{L})}e^{-k T_{\beta}}\left ( \frac{q^{-2j_{L}k}+\mathellipsis+q^{+2j_{L}k}}{k(q^{k/2}-q^{-k/2})^{2}}\right ),\qquad q=e^{-g_{s}}
\ee
in terms of $N_{\beta}^{(j_{L})}=\sum_{j_{R}}(-1)^{2j_{R}}(2j_{R}+1)N_{\beta}^{(j_{L},j_{R})}$, not in terms of the individual degeneracies. This is because the individual degeneracies may change as we vary the complex structure due to pairing of different $j_{R}$ into multiplets. There exists an extension of the topological string theory due to the advances in the microscopic derivation of the prepotential of the supersymmetric gauge theories \cite{Moore:1997dj,Nekrasov:2002qd,Nekrasov:2003rj}. This extension is equivalent turning on both the self and anti-self dual graviphoton fields \cite{Hollowood:2003cv}. The free energy turns out to be written in terms of the individual degeneracies     
\begin{align}\nonumber
&F(\omega;t,q)=\\
&\sum_{\beta\in H_{2}(X,{\mathbb Z})}\sum_{n=1}^{\infty}\sum_{j_{L},j_{R}}\frac{(-1)^{2j_{L}+2j_{R}}N_{\beta}^{(j_{L},j_{R})}\Big ((t\,q)^{-nj_{L}}+\mathellipsis+ (t\,q)^{nj_{L}} \Big )\Big ( (\frac{t}{q})^{-nj_{R}}+\mathellipsis+(\frac{t}{q})^{nj_{R}}\Big )}{n(t^{n/2}-t^{-n/2})(q^{n/2}-q^{-n/2})}e^{-nT_{\beta}}
\end{align}
where $q=e^{F_{+}}$ and $t=e^{F_{-}}$.
\medskip

\par{Our main interest is ${\cal N}=2$ supersymmetric $SU(N)$ gauge theories with or without matter. This type of theories are engineered using the toric geometries. The toric geometries are encoded in 1$d$ tri-valent graphs which show the degeneration loci of the toric action on them in terms of the slopes of the lines in the toric diagram. The topological vertex formalism offers an all genus computation of the topological string partition function for the toric Calabi-Yau threefolds.  }
\begin{figure}[h]
\begin{center}
\includegraphics[scale=0.2]{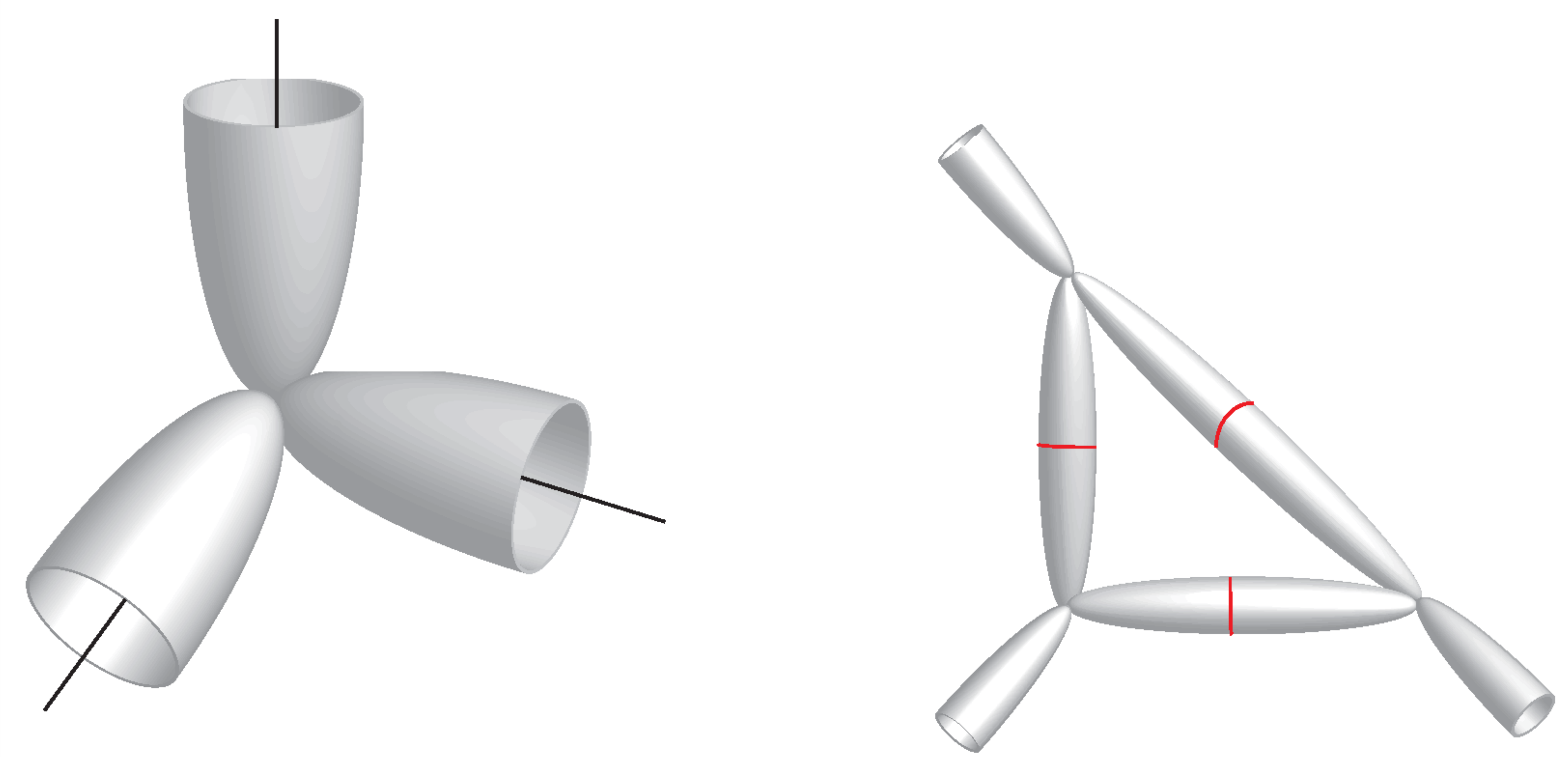}
\end{center}
\caption{The ${\mathbb C}^{3}$ patches. The patches can be glued to give rise to general three-fold, \textit{e.g.}, ${\cal O}(-1)\rightarrow {\mathbb P}^{2}$.\label{vertex2}}
\end{figure}
\par{The idea behind the topological vertex \cite{Iqbal:2002we,Aganagic:2003db} is to divide the geometry in ${\mathbb C}^{3}$ patches, as shown in figure \ref{vertex2}, and compute the open amplitudes for the individual patches. Different type patches arise in cutting the geometry but they all are related to each other by symmetry transformations of the theory. This allows to choose one generic patch and define the topological vertex for it. From the point of view of string theory this cutting is realized by placing Lagrangian branes/anti-branes. The topological vertex $C_{\lambda\mu\nu}(q)$ is defined through the open amplitude of ${\mathbb C}^{3}$
\be
Z(V_{1},V_{2},V_{3})=\sum_{\lambda,\mu,\nu}C_{\lambda\mu\nu}(q)\,\mbox{tr}_{\lambda}V_{1}\,\mbox{tr}_{\mu}V_{2}\,\mbox{tr}_{\nu}V_{3},
\ee
as a function of the parameter $q=e^{-g_{s}}$ with $g_{s}=\langle F_{+} \rangle$ being the topological string coupling constant. $V_{i}$'s are the holonomies supported on the branes, and the sums are over irreducible representation of $U(\infty)$. A suitable gluing algorithm joins the amplitudes of the individual patches into the amplitude of the whole geometry.  }\medskip
\par{ A combinatorial interpretation the topological vertex is given in terms of plane partitions, it is often called the ``crystal model" \cite{ORV, Takasaki}. It is identical, up to the MacMahon function, to the generating function of plane partitions subject to boundary  conditions dependent on the representations labeling its legs, see figure \ref{crystal}. The (formal) counting parameter on the combinatorics side is identified with the exponential of the topological string coupling constant. The generating function is computed by dividing the plane partition into slices, see \eqref{defslices}.}
\begin{figure}[h]
$\begin{array}{c@{\hspace{1in}}c} \multicolumn{1}{l}{\mbox{}} &
    \multicolumn{1}{l}{\mbox{}} \\ [-0.53cm]
 \includegraphics[width=2.5in]{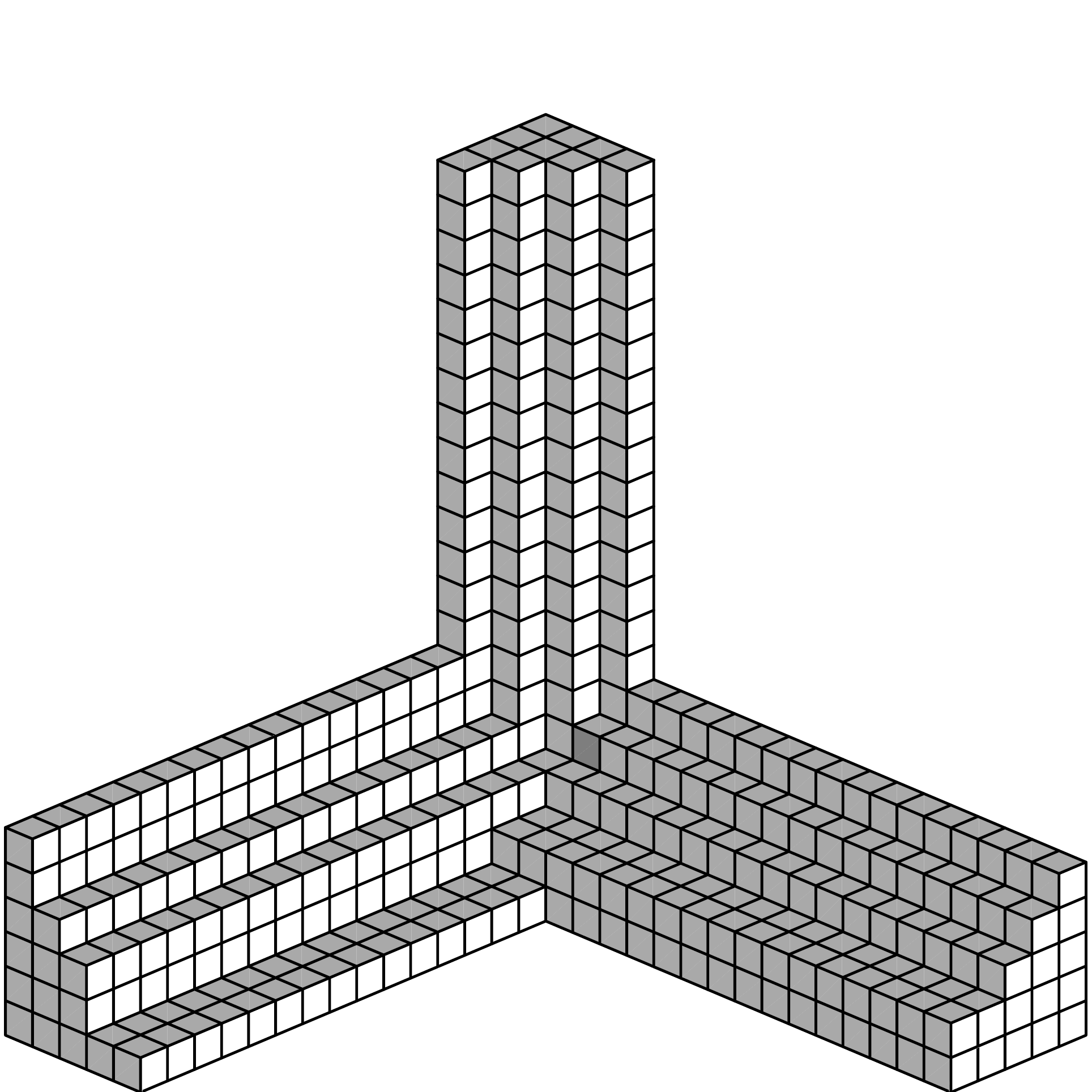} &
\includegraphics[width=2.5in]{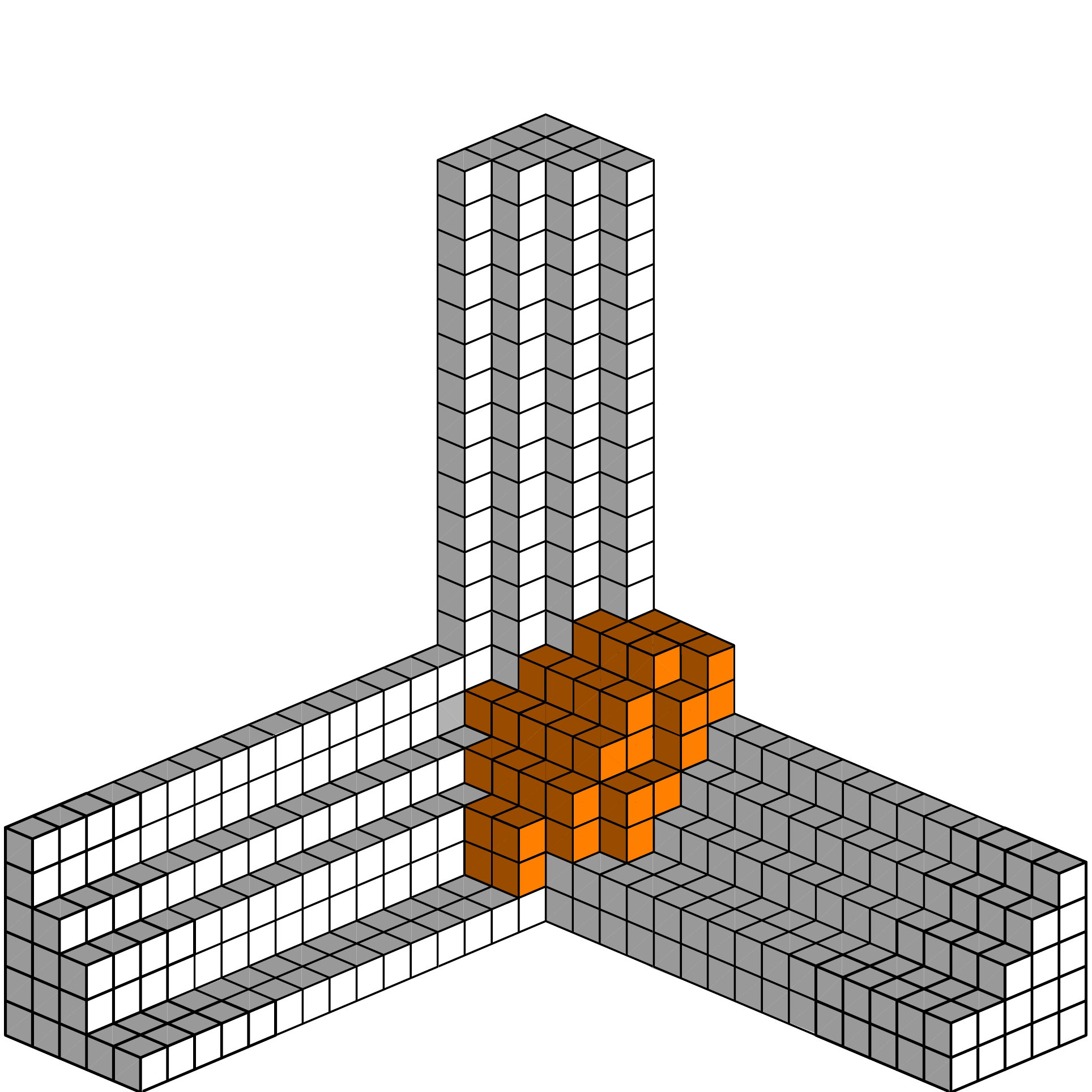}\\
\\ [0.4cm] \mbox{(a)} & \mbox{(b)}
\end{array}$
\caption{(a) $\pi_{\bullet}(\lambda,\mu,\nu)$ for
$\lambda=(6,4,3,1,1),\,\mu=(5,4,3,2,2),\, \nu=(4,3,2,1)$.  (b) An example
of $\pi(\lambda,\mu,\nu)$. The topological vertex is a generating function counting configurations $\pi(\lambda,\mu,\nu)$ for given $\lambda,\mu,\nu$, with a weight depending on slices of the plane partition.  \label{crystal}}
\end{figure}

\par{The refined topological vertex is originally constructed modifying this combinatorial interpretation of the usual topological vertex to reproduce the equivariant instanton part of the partition function. For the refined topological vertex, the direction along the slices of the plane partition plays a significant role unlike the usual one. The representation along this so-called preferred direction determines the weights of the slices, \textit{i.e.}, which slices are counted with $q$ or $t$. This is depicted in figure \ref{choice}. }

\begin{figure}[h]
\begin{center}
\includegraphics[scale=1]{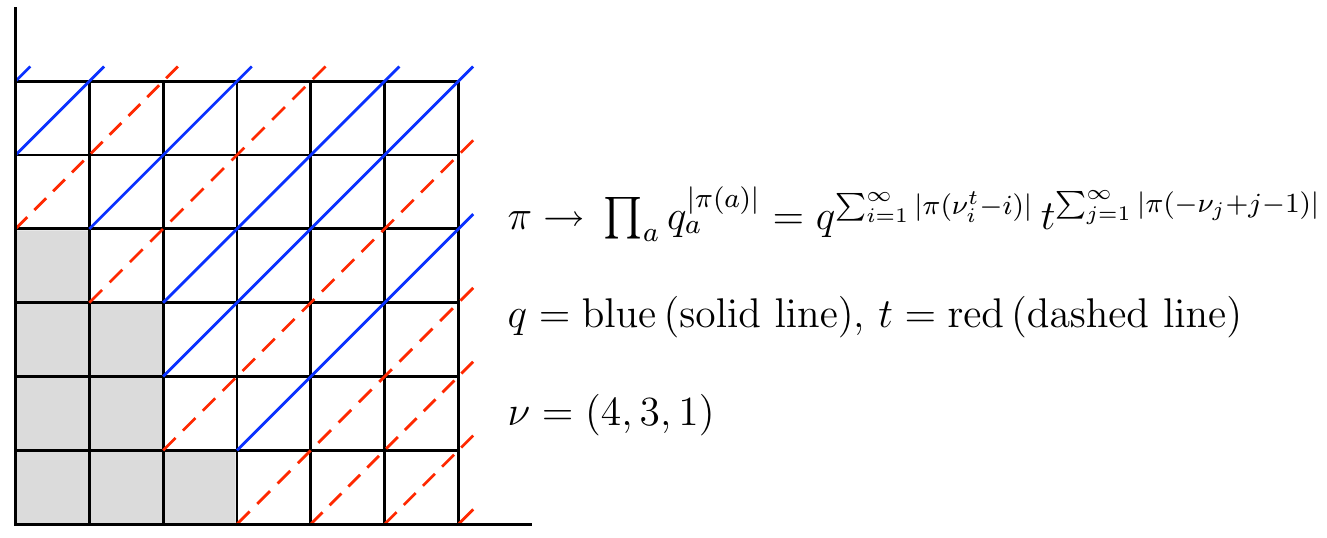}
\end{center}
\caption{The plane partition seen from above. The grey zone is the partition $\lambda$. Slices at times $s\in S_-(\lambda)$ are weighted by $t^{|\eta(s)|}$, and slices at times $s\in S_+(\lambda)$ are weighted by $q^{|\eta(s)|}$.\label{choice}}
\end{figure}
Consider a plane partition $\pi=\{\eta(s)\}$, and its slices $\eta(s)$.
The refined topological vertex is:
\be\label{eqtopvertexpartitions}
Z_{\lambda,\mu,\nu}(t,q) = \sum_{\pi,\,\delta\pi=(\lambda,\mu,\nu)}\,\, \prod_{s\in S_-(\lambda)} t^{\sum_i \eta_i(s)}\,\,\,\prod_{s\in S_+(\lambda)} q^{\sum_i \eta_i(s)}.
\ee

It is well known that if we have no boundary $\lambda=\mu=\nu=\emptyset$, and $t=q$, this is the Mac-Mahon function
\be
Z_{\emptyset,\emptyset,\emptyset}(q,q) = \prod_{k=1}^\infty (1-q^k)^{-k}.
\ee


\subsection{Weight of a plane partition}

The bounded refined topological vertex, is the generating function counting plane partitions
\be
Z = \sum_\pi\,\, {\cal W}(\pi)
\ee
 with the weight \eqref{eqtopvertexpartitions}:
\be
{\cal W}(\pi) = \prod_{s\in S_-(\lambda)}\, t^{\sum_{i=1}^N h_i(s)}\,\,\,\,\prod_{s\in S_+(\lambda)}\, q^{\sum_{i=1}^N h_i(s)}.
\ee
Notice that the restriction $h_N(s)\geq r_\lambda(s)$ can also be encoded by assigning a vanishing weight to plane partitions which do not satisfy that condition, and thus we write
\be\label{weightpplaneprodw}
{\cal W}(\pi) = \prod_{s=s_{\rm min}}^{s=s_{\rm max}} \prod_{i=1}^N\,w(h_i(s),s),
\ee
where
\be
w(h,s) = \left\{\begin{array}{ll}
t^h, & {\rm if}\,\, s\in S_-(\lambda) \,\,\,{\rm and}\,\,h-r_{\lambda}(s)\in \mathbb N\cr
q^h, & {\rm if}\,\, s\in S_+(\lambda) \,\,\,{\rm and}\,\,h-r_{\lambda}(s)\in \mathbb N \cr
0, & {\rm if}\,\, h-r_{\lambda}(s)\notin \mathbb N. \cr
\end{array}
\right.
\ee

More generally, one could generalize this problem, and require that the possible paths are restricted to a certain domain ${\cal D}$ in the $(h,s)$ plane, of an arbitrary shape (not necessarily given by a partition $\lambda$), or
even more generally, the paths could be weighted by an arbitrary ``landscape" potential  in the $(h,s)$ plane, \textit{i.e.}, one would like to give an arbitrary local weight $w(h,s)$, so that the weight of the path is the product \eqref{weightpplaneprodw}
and where the weight function $w(h,s)$ can be chosen at our convenience, for example a forbidden position is such that $w(h,s)=0$.
\medskip

In our case, for the refined topological vertex, non--forbidden positions will be chosen with weight $w(h,s)=t^h$ or $q^h$ depending on wether $s$ is in the ascending or descending side of the upper partition $\lambda$, \textit{i.e.}, if $w(h,s)\neq 0$
\be
w(h,s)=\left\{\begin{array}{l}
t^h, \quad {\rm when}\quad s\in S_-(\lambda)\cr
q^h, \quad {\rm when}\quad  s\in S_+(\lambda)
\end{array}\right. .
\ee

Notice that there is no need to assign values to $w(h,s)$ at positions $(h,s)$ which can never be visited by the paths (notion of shadow of a domain in \cite{eynard1}), and therefore we may also choose $w(h,s)=t^h$ or $q^h$ in a forbidden region which will never be visited.
\medskip

This implies that we need to choose a non-trivial weight, only for $s$ being an outer corner of $\lambda$. We write:
\be
S_-(\lambda) = \mathop{{\cup}}_{i=1}^{{\cal N}} ]u_{i-1},v_i]
\quad , \quad
S_+(\lambda) = \mathop{{\cup}}_{i=1}^{{\cal N}} ]v_i,u_i],
\ee
\textit{i.e.}, $v_i$ are the inner corners of $\lambda$ and $u_i$ are the outer corners of $\lambda$, see figure \ref{sl}.
Eventually, for the refined vertex we shall use:
\be
w(h,s)=\left\{\begin{array}{l}
t^h, \quad {\rm when}\quad u_{i-1}<s\leq v_i \cr
q^h, \quad {\rm when}\quad v_{i}<s< u_i \cr
q^h\,\,\left(1- \sum_{k=1}^{r_\lambda(u_i)-\frac{1}{2}|u_i|} \frac{\sin{\pi(k+h-r_\lambda(u_i))}}{\pi(k+h-r_\lambda(u_i))}\right)\,f(h-r_\lambda(u_i)), \quad {\rm when}\quad s=u_i
\end{array}\right.
\ee
where $f(h)$ can be any function of $h$ whose value is $1$ when $h\in \mathbb Z$.
We shall write
\be
w(h,s)=\re^{-V_s(h)},
\ee
and we call $V_s(h)$ the ``potential" at time $s$.

\setcounter{equation}{0}
\section{The matrix model for the refined topological vertex}

In this section, first, we want to review the general construction of the matrix model for plane partitions with different weights along each 2$d$ slice when we slice it. Later, the resulting matrix model is restricted based on the choices made for the refined topological vertex. It turns out that the potentials along most of the slices, more precisely, every slice but the first, last and along the outer corners, are linear. The simplicity of the potentials allows us to group slices into building blocks between the slices along the outer corners which we can glue afterwards. Within the individual blocks all matrix integrals but one can be performed leading to a significant simplification of the model. 

\subsection{The matrix model for the plane partitions}
In \cite{eynard1}, it has been proved that the sum over plane partitions with arbitrary weights $w(h,s)$ can be rewritten as a matrix integral.
Namely, theorem 3.1 in \cite{eynard1}, says that the generating function of plane partitions is proportional to the following multi-matrix integral
\begin{align}\label{Zmatrixmodel}
{\cal Z}&=\int_{(H_{N})^{s_{\mbox{\tiny max}}-s_{\mbox{\tiny min}}-1 }}\prod_{s=s_{\mbox{\tiny min}}+1}^{s_{\mbox{\tiny max}}-1}dM_{s}\int_{(i\,H_{N})^{s_{\mbox{\tiny max}}-s_{\mbox{\tiny min}}}}\prod_{s'=s_{\mbox{\tiny min}}+\frac{1}{2}}^{s_{\mbox{\tiny max}}-\frac{1}{2}}dR_{s'} \prod_{s=s_{\mbox{\tiny min}}}^{s_{\mbox{\tiny max}}}e^{-\mbox{\scriptsize Tr}\, V_{s}(M_{s})}\,\\ 
&\times \prod_{s'=s_{\mbox{\tiny min}}+\frac{1}{2}}^{s_{\mbox{\tiny max}}-\frac{1}{2}}e^{-\mbox{\scriptsize Tr}\, U_{s'}(R_{s'})}\prod_{s'=s_{\mbox{\tiny min}}+\frac{1}{2}}^{s_{\mbox{\tiny max}}-\frac{1}{2}}e^{\mbox{\scriptsize Tr}\, R_{s'}(M_{s'+\frac{1}{2}}-M_{s'-\frac{1}{2}})},
\end{align}
where the potentials $V_s(h)$ encode the weights $w(h,s)=\re^{-V_s(h)}$ and the potential $U_{s'}(r)$ encodes the jumps by $\pm 1/2$, \textit{i.e.}, $e^{-U_{s'}(r)}$ is the Fourrier--transform of $\delta(h-\frac 1 2)+\delta(h+\frac 1 2)$, namely it is
\be
\re^{-U_{s'}(r)} = \re^{-i\pi r}+\re^{i\pi r} = 2\cos{\pi r}.
\ee
The boundary conditions at the slices $-s_{\rm min}$ and $s_{\rm max}$ are encoded in the following fixed diagonal matrixes which are not integrated over
\begin{align}
&M_{s_{\rm min}}={\rm diag}(h_1(s_{\rm min}),h_2(s_{\rm min}),\dots,h_N(s_{\rm min})),\\
&M_{s_{\rm max}}={\rm diag}(h_1(s_{\rm max}),h_2(s_{\rm max}),\dots,h_N(s_{\rm max})).
\end{align}

The proof of that formula is given in \cite{eynard1}, but let us just sketch it here. 
It uses the fact that the generating function of non--intersecting paths is a determinant (the Lindstroem--Gessel--Viennot determinant LGV \cite{LGV})
\be
{\cal Z} = \det_{1\leq i,j\leq N}\,\, {\cal W}_{i,j}
\ee
where ${\cal W}_{i,j}$ is the weight of an individual path 
going from $h_i(s_{\rm min})$ to $h_j(s_{\rm max})$, \textit{i.e.},
\be
\ba
{\cal W}_{i,j} =  \int_{-\infty}^\infty \prod_{s=s_{\rm min}}^{s_{\rm max}} \, w(x_s,s)\,dx_s\,\quad\,\delta(x_{s_{\rm min}}-h_i(s_{\rm min}))\,\delta(x_{s_{\rm max}}-h_j(s_{\rm max}))\,\cr
\times\prod_{s=s_{\rm min}}^{s_{\rm max}-1}\,\, \left(\delta(x_{s+1}-x_s-\frac 1 2)+\delta(x_{s+1}-x_s+\frac 1 2)\right).
\ea
\ee
The $\delta$-functions appearing in the weights can be Fourier transformed
\be
\delta \left (x_{s+1}-x_s-\frac{1}{2} \right ) +\,\delta \left (x_{s+1}-x_s+\frac{1}{2} \right )=\int_{-\infty}^{\infty} dr \,e^{2i\pi\, r (x_{s+1}-x_s)} \left [ e^{-i\pi\, r} + e^{i\pi\, r} \right ].
\ee
Inserting the Fourier transform into the LGV determinant yields (after symmetrization over the integration variables) the determinants $\det \re^{2i\pi\,r_i h_j}$, which we recognize as the Harish--Chandra Itzykson--Zuber integral \cite{HC,IZ}:
\be\label{eqHCIZ}
\det\left(\re^{2i\pi\,\,r_i h_j}\right) = \Delta(h)\,\Delta(r)\, \int_{U(N)}\, dU\,\re^{2i\pi\,\,\tr \, h U r U^\dagger},
\ee
where $h={\rm diag}(h_1,\dots,h_N)$, $r={\rm diag}(r_1,\dots,r_N)$ and $\Delta(h)=\prod_{i>j} (h_i-h_j)$, $\Delta(r)=\prod_{i>j} (r_i-r_j)$ are the Vandermonde determinants.
This allows to rewrite the sum over self--avoiding paths as a matrix integral. The eigenvalues of the matrices are the $h_{i}$'s and the $r_{i}$'s for each time $s$. More precisely, $M_s$ is a matrix whose eigenvalues are the $h_i(s)$, and $R_{s+\frac 1 2}$ is a matrix whose eigenvalues are the Fourier variables corresponding to the delta functions encoding the jumps from $s\to s+1$. The angular parts of those matrices are provided by the $U\in U(N)$ in the Harish-Chandra Itzykson--Zuber integrals \eqref{eqHCIZ}.
\medskip

We shall not enter further details of the derivation of formula \eqref{Zmatrixmodel} here, since we are only interested in an application of this general formula, and thus we refer the interested reader to \cite{eynard1} for the full derivation.
Another presentation of the derivation is also explained in \cite{OSY}.

\subsection{Integrating out the intermediate matrices}

The matrix model \eqref{Zmatrixmodel} is a multi--matrix integral, known as the ``chain of matrices" (see \cite{DGZ}). This is an integrable, exactly solvable matrix model (see \cite{DGZ, Mehtabook}). However, in our case, it can be further simplified. Indeed, for most values of $s$ (\textit{i.e.}, $s\neq u_i$), the potentials $V_s(M_s)$ are of the form 
\be
V_s(M_s)=-M_s\,\ln q
\qquad {\rm or}\qquad
V_s(M_s)=-M_s\,\ln t,
\ee
\textit{i.e.}, the potentials are linear in $M_s$. Most of the integrals over $M_s$ can be performed explicitly, and our multi--matrix model reduces to a matrix model with fewer matrices. The only $M_s$ which can not be integrated easily, are those corresponding to values of $s$ where the potential $V_s(M_s)$ is not linear,that is when $s=u_i$ is at an outer corner of the upper partition $\lambda$.

\begin{figure}[h]
\begin{center}
\includegraphics[scale=0.5]{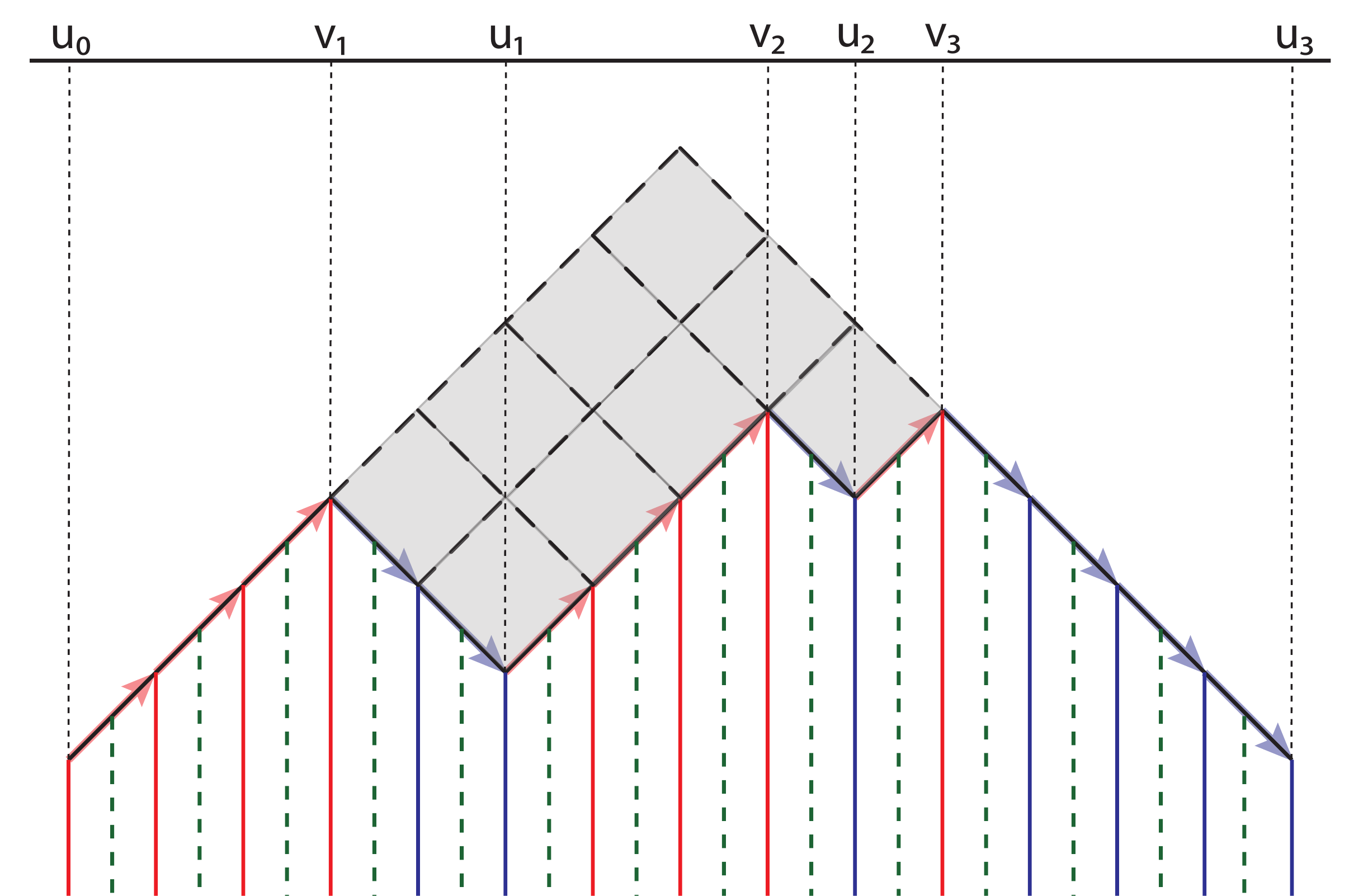}
\end{center}
\caption{The choice of weights. Red slices (i.e. ascending edges of $\lambda$) have weight $t$, and blue slices have weight $q$.\label{sl}}
\end{figure}

\medskip

Therefore, for any given partition $\lambda$ 
we will group slices from one outer corner to the next one into a basic building block. Let us call the block from the $(i-1)^{\rm th}$ outer slice to the $i^{th}$ as 
\begin{align}\nonumber
&{\cal Z}_{i-1,i}(M_{u_{i-1}},R_{v_i+\frac 1 2},M_{u_i}) 
\equiv\int_{(H_{N})^{(u_{i}-u_{i-1}-1)}}\prod_{s=u_{i-1}+1}^{u_{i}-1}dM_{s}\int_{(i\,H_{N})^{(u_{i}-u_{i-1}-1)}}\prod_{\substack{s'=u_{i-1}+1/2\\s'\neq v_{i}+1/2}}^{u_{i}-1}dR_{s'}\\ 
&\times\prod_{s=u_{i-1}+1}^{u_{i}-1}e^{-\mbox{\scriptsize Tr}\, V_{s}(M_{s})}\prod_{s'=u_{i-1}+1/2}^{u_{i}-1/2}e^{-\mbox{\scriptsize Tr}\, U_{s'}(R_{s'})}\prod_{s'=u_{i-1}+1/2}^{u_{i}-1/2}e^{\mbox{\scriptsize Tr}\, R_{s'}(M_{s'+1/2}-M_{s'-1/2})}.
\end{align}

The appropriate weight to compute the refined partition function is encoded in the potentials $V_{s}(M_{s})$:
\begin{equation*}
V_{s}(M_{s})=\left \{\begin{array}{ll}\nonumber -\ln t\, M_{s},& \mbox{for}\,\,\,u_{i-1}<s \leq v_{i} \\ \nonumber -\ln q\, M_{s},&\mbox{for}\,\,\,v_{i}< s<u_{i}.  \end{array}  \right.
\end{equation*}
The integrals along the slices between the outer corners can be performed giving rise to matrix $\delta$-functions
\begin{align}\nonumber
&\int_{(H_{N})^{(u_{i}-u_{i-1}-1)}}\prod_{s=u_{i-1}+1}^{u_{i}-1}dM_{s}\prod_{s=u_{i-1}+1}^{u_{i}-1}e^{-\mbox{\scriptsize Tr}\, V_{s}(M_{s})}\prod_{s'=u_{i-1}+1/2}^{u_{i}-1/2}e^{\mbox{\scriptsize Tr}\, R_{s'}(M_{s'+1/2}-M_{s'-1/2})}\\
&=\prod_{s'=u_{i-1}+1/2}^{v_{i}-1/2}\delta(R_{s'}-R_{s'+1}+\ln t)\prod_{s'=v_{i}+1/2}^{u_{i}-3/2}\delta(R_{s'}-R_{s'+1}+\ln q)\, e^{\mbox{\scriptsize Tr}\, ( M_{u_{i}}R_{u_{i}-1/2}-M_{u_{i-1}}R_{u_{i-1}+1/2})}
\end{align}
After performing each $R_{s'}$ integration but $R_{v_{i}+1/2}$ in the interval $]u_{i-1},u_{i}[$ one obtains 
\begin{align}\nonumber
{\cal Z}_{i-1,i}(M_{u_{i-1}},R,M_{u_i})&=e^{\mbox{\scriptsize Tr}\,R(M_{u_{i}}-M_{u_{i-1}})}e^{(v_{i}-u_{i-1})\ln t\, \mbox{\scriptsize Tr}\, M_{u_{i-1}}+(u_{i}-v_{i}-1)\ln q \,\mbox{\scriptsize Tr}\, M_{u_{i}}}\\ 
&\times \prod_{k=1}^{v_{i}-u_{i-1}}e^{-\mbox{\scriptsize Tr}\, U(R-k\ln t)}\prod_{k=1}^{u_{i}-v_{i}}e^{-\mbox{\scriptsize Tr}\, U( R+(k-1)\ln q)}
\end{align}
The last two products can alternatively be written as
\begin{eqnarray}\nonumber
 \prod_{k=1}^{v_{i}-u_{i-1}}e^{-\mbox{\scriptsize Tr}\, U(R-k\ln t)}&\propto&e^{(v_{i}-u_{i-1})\mbox{\scriptsize Tr}\,R/2} \frac{\det \left(-t\, e^{-R};t \right)_{\infty}}{\det \left(-t^{v_{i}-u_{i-1}+1}\, e^{-R};t \right)_{\infty}}\\ \nonumber
 \prod_{k=1}^{u_{i}-v_{i}}e^{-\mbox{\scriptsize Tr}\, U( R+(k-1)\ln q)}& \propto&e^{-(u_{i}-v_{i})\mbox{\scriptsize Tr}\,R/2}\frac{\det \left(-e^{R};q\right)_{\infty}}{\det \left(-q^{(u_{i}-v_{i})}\, e^{R};q \right)_{\infty}}
\end{eqnarray}
up to factors independent on $R$. $(a;q)_{\infty}$ is the Pochammer symbol and is defined as
\begin{equation}
(a;q)_{\infty}\equiv\prod_{k=0}^{\infty}(1-a\,q^{k}).
\end{equation}
\par{The individual blocks thus take the form}
\begin{align}\label{eqZi-1i}
{\cal Z}_{i-1,i}&=e^{\mbox{\scriptsize Tr}\,R_{v_{i}+1/2}(M_{u_{i}}-M_{u_{i-1}})}e^{ (v_{i}-u_{i-1})\ln t\, \mbox{\scriptsize Tr}\, M_{u_{i-1}}+(u_{i}-v_{i}-1)\ln q \,\mbox{\scriptsize Tr}\, M_{u_{i}}}e^{(2v_{i}-u_{i}-u_{i-1})\mbox{\scriptsize Tr}\,R_{v_{i}+1/2}/2}\\ 
&\times  \frac{\det \left(-t\, e^{-R_{v_{i}+1/2}};t \right)_{\infty}}{\det \left(-t^{v_{i}-u_{i-1}+1}\, e^{-R_{v_{i}+1/2}};t \right)_{\infty}}\frac{\det \left(-e^{R_{v_{i}+1/2}};q\right)_{\infty}}{\det \left(-q^{(u_{i}-v_{i})}\, e^{R_{v_{i}+1/2}};q \right)_{\infty}}.
\end{align}

The whole integral is then given in terms of these blocks, hence, we end up with the following chain matrix model for the refined topological vertex
\begin{equation}\label{matrixmodel}
{\cal Z}=\int_{(H_{N})^{{\cal N}}}\prod_{i=1}^{{\cal N}}dM_{u_{i}}\int_{(i H_{N})^{{\cal N}}}dR_{v_{i}+1/2}\,e^{-\mbox{\scriptsize Tr}\, V_{u_{i}}(M_{u_{i}})}\,{\cal Z}_{i-1,i}(M_{u_{i-1}},R_{v_i+\frac 1 2},M_{u_{i}}).
\end{equation}
Let us remind that the representations of the refined topological vertex along the unpreferred directions are imposed by the fixed matrices at the first and last slices. The representation can be alternatively given using the inner and outer coordinates which appear in the above matrix integrals.


\section{The matrix model for the refined strip geometry}\label{toric}
\par{In this section we want to extend the matrix model for the refined vertex to more general cases and propose a crystal model for the refined strip geometry \cite{Iqbal:2004ne,Taki:2007dh}. Two pieces of these geometries can be glued along the external legs to engineer interesting $SU(N)$ geometries with extended supersymmetry, including 4$d$ superconformal theories.} 
\par{Our model will resemble the bubbling picture of \cite{Gomis:2007kz}. We will excise a region such as the one in figure \ref{young} whose sizes are related to the K\"{a}hler parameters in the strip geometry. We grow the crystal in the remaining unbounded region. This geometry requires to redefine the weights on the slices along the outer corners to get a combinatorial model, each such slice is counted with $t$ instead of $q$ like in the case of the refined topological vertex. We want to work out the example shown in figure \ref{young} in detail. From our approach it will be clear that there is no obstruction in generalizing this example to a longer strip.} 

\begin{figure}[h]
\begin{center}
\includegraphics[scale=0.5]{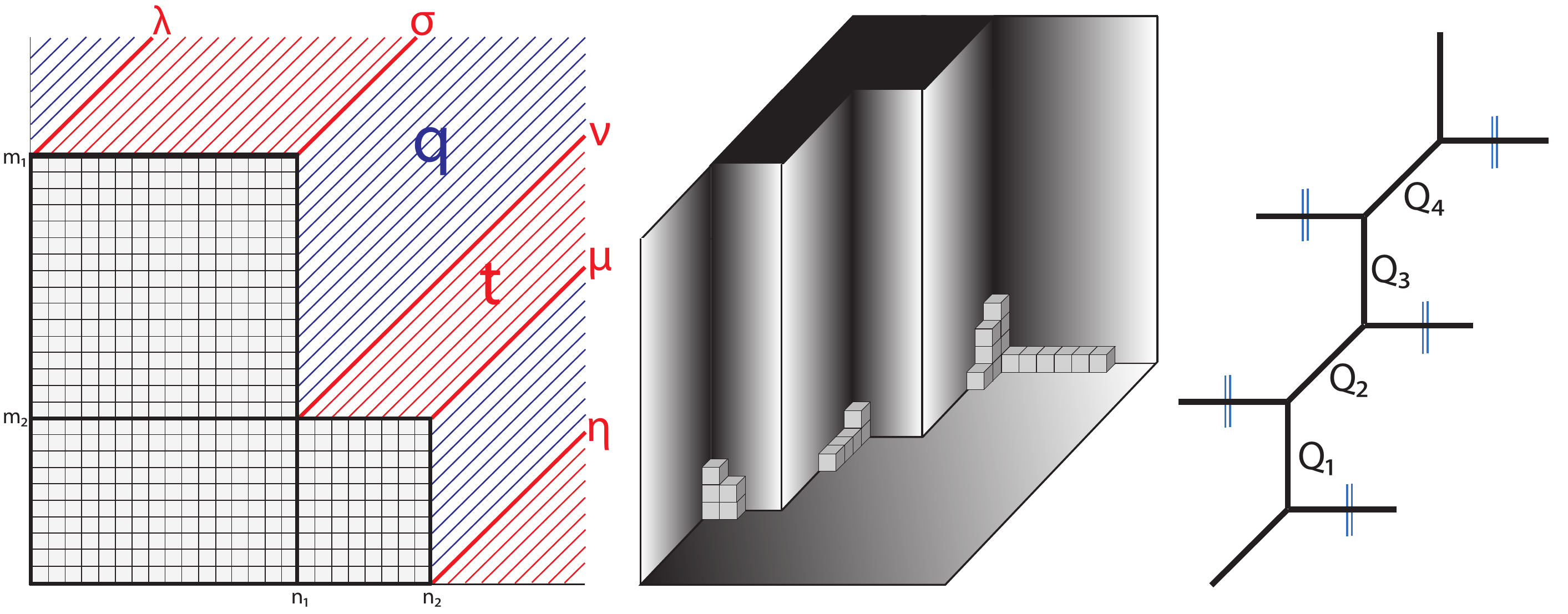}
\end{center}
\caption{(a) the definition of slices, (b) the region where the crystal grows, (c) the refined strip geometry}
\label{young}
\end{figure}

\par{We will compute the generating function for the crystal using the transfer matrix approach. The following identities are crucial for our computation. The vertex operators $\Gamma_{\pm}(1)$ and the Hamiltonian $L_{0}$ satisfy}

\be
\Gamma_{-}(1)|\mu\rangle=\sum_{\nu\succ\mu}|\nu\rangle,\qquad \Gamma_{+}(1)|\mu\rangle=\sum_{\nu\prec\mu}|\nu\rangle, \qquad\mbox{and}\,\qquad q^{L_{0}}|\mu\rangle=q^{|\mu|}|\mu\rangle.
\ee
\par{We will compute the generating function in the following way: we will assign an arbitrary Young diagram to the slices along the inner and outer edges and find the contributions with these `boundary' conditions. In other words we are inserting identity operators at certain places in the generating function to ease the computation. Later we sum over all possible Young diagrams along these slices and match these sums to the summations over Young diagram we obtain from the refined topological vertex gluing. It will be obvious that the continuation of  this procedure iteratively and straightforward. The generating function for the example given in the figure \ref{young} is given by}

\begin{align}\nn
Z&=\sum_{\lambda,\sigma,\nu,\mu,\eta}\langle\emptyset|\prod_{i=1}^{\infty}q^{L_{0}}\Gamma_{+}(1)|\lambda\rangle\langle\lambda|t^{L_{0}}\Gamma_{-}(1)\prod_{i=1}^{n_{1}-1} t^{L_{0}}\Gamma_{-}(1)|\sigma\rangle\langle\sigma|t^{L_{0}}\Gamma_{+}(1)\prod_{i=1}^{m_{1}-m_{2}-1}q^{L_{0}}\Gamma_{+}(1)|\nu\rangle\\
&\times\langle\nu|t^{L_{0}}\Gamma_{-}(1)\prod_{i=1}^{n_{2}-n_{1}-1}t^{L_{0}}\Gamma_{-}(1)|\mu\rangle\langle\mu|t^{L_{0}}\Gamma_{+}(1)|\eta\rangle\langle\eta|t^{L_{0}}\prod_{i=1}^{\infty}\Gamma_{-}(1)t^{L_{0}}|\emptyset\rangle
\end{align}
Note that the skew Schur functions can be written as
\be
s_{\lambda/\mu}(x_{1},x_{2},\mathellipsis,x_{n},0,0,\mathellipsis)=\sum_{\lambda\succ\lambda(1)\succ\mathellipsis\succ\lambda(n-1)\succ\mu}x_{1}^{|\lambda|-|\lambda(1)|}x_{2}^{|\lambda(1)|-|\lambda(2)|}\mathellipsis x_{n}^{|\lambda(n-1)|-|\mu|}.
\ee

Using the above identity and others collected in the appendix after some algebraic manipulations the generating functions takes the following form
\begin{align}\nn
Z&=M(t,q)^{3}\sum_{\sigma,\mu,\tau,\{\alpha_{i}\}}(-q^{m_{2}})^{|\alpha_{1}|}\,(-t^{n_{2}-n_{1}})^{|\alpha_{2}|}\,(-q^{m_{1}-m_{2}})^{|\alpha_{3}|}\,(-t^{n_{1}})^{|\alpha_{4}|}\,t^{|\alpha_{1}|+|\alpha_{2}|+|\alpha_{3}|+|\alpha_{4}|}\,\\ \nn
& \times q^{-|\mu|-|\sigma|}t^{-|\tau|}
\,\,\,\, s_{\alpha_{1}}(1,t,\mathellipsis) s_{\alpha_{1}^{t}/\mu}(1,q,\mathellipsis)s_{\alpha_{2}^{t}/\mu}(1,t,\mathellipsis)s_{\alpha_{2}/\tau}(1,q,\mathellipsis)s_{\alpha_{3}/\tau}(1,t,\mathellipsis) \\
& \times s_{\alpha_{3}^{t}/\sigma}(1,q,\mathellipsis) s_{\alpha_{4}^{t}/\sigma}(1,t,\mathellipsis)s_{\alpha_{4}}(1,q,\mathellipsis).
\end{align}

This expression looks similar to the sums we obtain when we glue refined vertices with the choice of the preferred direction along the external legs. Let us briefly review the refined topological vertex computation. The partition function is given as 

\begin{align}\nn
Z&=\sum_{\mu,\nu,\lambda,\tau}(-Q_{1})^{|\mu|}(-Q_{2})^{|\nu|}(-Q_{3})^{|\lambda|}(-Q_{4})^{|\tau|}C_{\mu\emptyset\emptyset}(t,q)C_{\mu^{t}\nu\emptyset}(q,t)C_{\lambda\nu^{t}\emptyset}(t,q)C_{\lambda^{t}\tau\emptyset}(q,t)C_{\emptyset\tau^{t}\emptyset}(t,q)\\\nn
&=\sum_{\mu,\nu,\lambda,\tau,\{\eta_{i}\}}(-Q_{1}t^{1/2}q^{1/2})^{|\mu|}(-Q_{2}t^{1/2}q^{1/2})^{|\nu|}(-Q_{3}t^{1/2}q^{1/2})^{|\lambda|}(-Q_{4}t^{1/2}q^{1/2})^{|\tau|}t^{|\eta_{2}|}q^{-|\eta_{1}|-|\eta_{3}|}\\
&\times s_{\mu}(1,t,\mathellipsis)s_{\mu^{t}/\eta_{1}}(1,q,\mathellipsis)s_{\nu^{t}/\eta_{1}}(1,t,\mathellipsis)s_{\nu/\eta_{2}}(1,q,\mathellipsis)s_{\lambda/\eta_{2}}(1,t,\mathellipsis)s_{\lambda^{t}/\eta_{3}}(1,q,\mathellipsis)\\\nn
&\times s_{\tau^{t}/\eta_{3}}(1,t,\mathellipsis)s_{\tau}(1,q,\mathellipsis)
\end{align}

The refined vertex result matches the crystal model under the identification of the K\"{a}hler parameters 
\be
Q_{1}\sqrt{\frac{q}{t}}=q^{m_{2}},\qquad Q_{2}\sqrt{\frac{q}{t}}=t^{n_{2}-n_{1}},\qquad Q_{3}\sqrt{\frac{q}{t}}=q^{m_{1}-m_{2}},\qquad Q_{4}\sqrt{\frac{q}{t}}=t^{n_{1}}
\ee 
up to MacMachon fucntions
\be
Z_{vertex}=M(t,q)^{3}Z_{crystal}.
\ee

\par{Having found a crystal model for the refined strip geometry it is straight forward to derive the corresponding matrix model. As mentioned the counting scheme differs from the one of the refined topological vertex at the outer corner. Luckily this difference can be very easily accommodated in our previous computation. We integrated out the intermediate slices between two outer corners and defined the individual blocks ${\cal Z}_{i-1,i}$. Since there are no modifications for these intermediate slices the blocks remain unchanged. However, when we are computing the partition function we need to integrate with respect to these modified slices. Therefore we only need to change the potential along these slices   }

\begin{equation}
{\cal Z}=\int_{(H_{N})^{{\cal N}}}\prod_{i=1}^{{\cal N}}dM_{u_{i}}\int_{(i H_{N})^{{\cal N}}}dR_{v_{i}+1/2}\,e^{-\mbox{\scriptsize Tr}\, {\widetilde V}_{u_{i}}(M_{u_{i}})}\,{\cal Z}_{i-1,i}(M_{u_{i-1}},R_{v_i+\frac 1 2},M_{u_{i}}),
\end{equation}
with ${\widetilde V}_{u_{i}}(M_{u_{i}})=-\ln t \,M_{u_{i}}$.


\section{Spectral curve and limitting shape}

\par{Our next aim is to find the spectral curve of our matrix model. It was derived in \cite{eynchain, eynpratschain} that the spectral curve of a chain of matrices is obtained as the ``saddle point'' of the corresponding matrix model. Before we move to the details, we want to point out a caveat of the saddle point approach. Unless the potential is polynomial it is possible that there are infinitely many saddle points of the potential. The spectral curve is the one which gives the global minimum, hence, for such potentials it is almost impossible to check whether this condition is satisfied. The lack of this explicit verification of the global minimum is not a big hindrance to obtain the spectral curve. However, when $\beta$ is rational, all those conditions can be realized with algebraic functions, and the natural guess, is to look for the smallest possible degree algebraic curve which satisfies all the conditions.}
\medskip

\par{Let us first review the basic strategy to find the spectral curve using saddle point approach. Later we motivate and write down our algebraic ansatz. Finally, we comment on limiting shapes of the crystals. }

\subsection{Rules for finding the spectral curve}

According to  \cite{eynchain, eynpratschain}, the spectral curve is described as follows: there must exist a Riemann surface ${\cal C}$ (to be determined later from consistency conditions), and some analytical functions ${\cal C}\to \mathbb CP^1$ (one per matrix in the chain), which we note:
\begin{align}
z\mapsto x_s(z)&, \qquad s_{\rm min}\leq s\leq s_{\rm max}\\
z\mapsto r_{s'}(z)&, \qquad s_{\rm min}+\frac{1}{2}\leq s'\leq s_{\rm max}-\frac{1}{2}
\end{align}
such that the saddle point conditions read as
\begin{align}
r_{s-1/2}(z)-r_{s+1/2}(z)&= V'_s(x_s(z))\\
x_{s'+1/2}(z)-x_{s'-1/2}(z)&= U'_{s'}(r_{s'}(z)) = -\,\frac{1}{2}\,\,\,\frac{\re^{r_{s'}(z)}-1}{\re^{r_{s'}(z)}+1}
\end{align}

While deriving our matrix model we divided the times into blocks based on the representation along the preferred direction. Similarly, we will treat these regions separately: 
\medskip

$\bullet$ For slices in the region $u_{i-1}< s\leq v_i$ the potential is
\be
V_s(x) = -x\,\ln t.
\ee
This allows us to define a function $R_i(z)=r_{v_i+\frac 1 2}(z)$ such that
\be
\forall\,s\,,\,\,\, u_{i-1}\leq s \leq v_i\, , \qquad r_{s+1/2}(z) = R_i(z) + (s-v_i)\ln t,
\ee
and thus
\be
x_{s+1}(z)-x_{s}(z) =  -\,\frac{1}{2}\,\,\,\frac{t^{s-v_i}\,\re^{R_i(z)}-1}{t^{s-v_i}\,\re^{R_i(z)}+1}.
\ee
Let us define:
\be
Y_i(z)=\re^{R_i(z)},\qquad\mbox{and}
\qquad
X_s(z) = t^{-x_s(z)}.
\ee
Summing over some intermediate slices
\be
x_{v_i}(z)-x_{s}(z) = -\frac 1 2 \,\sum_{k=1}^{v_i-s} \frac{Y_i(z)-t^k}{Y_i(z)\,+t^k} = \frac{s-v_i}{2} + \sum_{k=1}^{v_i-s} \frac{t^k}{Y_i(z)\,+t^k}.
\ee
In the limit $\ln t\to 0$ this gives
\be
 \frac{X_{s}(z)}{X_{v_i}(z)} \sim \frac{t^{\frac{v_i-s}2}+ t^{\frac{s-v_i}2}\, Y_i(z)}{1+Y_i(z)}\,\, (1+ O(\ln t))
\ee

$\bullet$ Similarly, for the region $v_{i}< s< u_i$ we have
\be
V_s(x) = -x\,\ln q.
\ee
this implies that
\be
r_{s+1/2}(z) = R_i(z) + (s-v_i)\ln q
\ee
and thus
\be
x_{s+1}(z)-x_{s}(z) =  -\,\frac{1}{2}\,\,\,\frac{q^{s-v_i}\,\re^{R_i(z)}-1}{q^{s-v_i}\,\re^{R_i(z)}+1}.
\ee
In the same way we have
\be
x_{s}(z)-x_{v_i}(z) = -\,\frac 1 2 \,\sum_{k=0}^{s-v_i-1} \frac{q^kY_i(z)-1}{q^kY_i(z)+1}
= \,\frac {s-v_i} 2 - \,\sum_{k=0}^{s-v_i-1} \frac{q^kY_i(z)}{q^kY_i(z)+1}.
\ee
Let us define again
\be
X_s(z) = t^{-x_s(z)}.
\ee
In the limit $\ln t\to 0$ this gives
\be
\left(\frac{X_{s}(z)}{X_{v_i}(z)}\right)^{1/\beta} \sim \frac{q^{\frac{v_i-s}2}+ q^{\frac{s-v_i}2}\, Y_i(z)}{1+Y_i(z)}\,\, (1+ O(\ln t))
\ee
where we have introduced
\be
\beta = \ln t/\ln q
\ee
that is assumed to be $O(1)$ when $\ln t\to 0$.

$\bullet$ At times $s=u_i$, we have
\be
r_{u_i-\frac 1 2}(z)-r_{u_i+\frac 1 2}(z) = V'_{u_i}(x_{u_i}(z)) .
\ee
Remember that the potential $V_{u_i}(x)$ is in this case not well defined. The only requirement is that it is worth $V_{u_i}(x) = -x\,\ln q$ if $x$ is an allowed position, and $V_{u_i}(x)$ has a logarithmic singularity at all non-allowed positions.
This means that within the allowed domain, we can replace $V_{u_i}(x)$ by $-x\,\ln q$, and at the boundary of the allowed domain, it has a singularity of the form $\ln(X_{u_i}(z)-t^{r_\lambda(u_i)}) +O(\ln t)$.
This implies that
\be
\frac{Y_{i+1}(z)}{Y_i(z)} = t^{v_{i+1}-u_i}\,q^{u_i-v_i}
\ee
\be\label{eqzeroYi}
Y_i(z)=0, \qquad {\rm if}\,\, X_{u_i}(z) = t^{r_\lambda(u_i)}
\ee
\be\label{eqpoleYi}
Y_i(z)=\infty, \qquad {\rm if}\,\, X_{u_{i-1}}(z) = t^{r_\lambda(u_{i-1})}
\ee

\subsection{Ansatz: an algebraic spectral curve}\label{sectionspcurvealg}

We see that to leading order in $\ln t$, we need to find functions $Y_i(z)$ which are analytical on a certain Riemann surface ${\cal C}$, which have a zero at
$X_{u_i}(z)=t^{r_\lambda(u_i)}$ and a pole at $X_{u_{i-1}}(z)=t^{r_\lambda(u_{i-1})}$.
Moreover, we need to find functions $X_{v_i}(z)$ analytical on ${\cal C}$ which vanish at $Y_i(z)=-1$ with exponent $1$ and $\beta$, and which diverge at $Y_i=-t^{v_i-u_{i-1}}$ with exponent $1$ and 
 diverge at $Y_i=-q^{v_i-u_i}$ with exponent $\beta$, and such that
\be
\begin{array}{l}
X_s(z) = X_{v_i}(z)\,\,\frac{t^{\frac{v_i-s}{2}}+t^{\frac{s-v_i}{2}}Y_i(z)}{1+Y_i(z)}, \qquad\qquad u_{i-1}< s\leq v_i \cr
X_s(z) = X_{v_i}(z)\,\,\left(\frac{q^{\frac{v_i-s}{2}}+q^{\frac{s-v_i}{2}}Y_i(z)}{1+Y_i(z)}\right)^\beta, \qquad v_i\leq s < u_i \cr
\cr
\frac{Y_{i+1}(z)}{Y_i(z)} = t^{v_{i+1}-u_i}\,q^{u_i-v_i},\qquad s=u_{i}.
\end{array}
\ee

\medskip

All those requirements are of algebraic nature, and it is natural to look for an algebraic curve ${\cal C}$ and algebraic functions of the minimal possible degrees satisfying all the requirements.

\smallskip
It is rather easy to find this minimal degree algebraic curve, as one can see in the examples of section \ref{ex} below. As we mentioned, there is no proof that this minimal degree algebraic curve, is indeed the spectral curve of the matrix model, \textit{i.e.}, that it is indeed the leading saddle point corresponding to our matrix integral, and that there is no other saddle point with lower energy. However, in many examples, this choice looks very reasonable. For the unrefined case $t=q$, this minimal degree algebraic curve coincides with the one found by Kenyon--Okounkov--Sheffield \cite{KOS}. See an explicit example in section \ref{ex} below.
\medskip

From the point of view of string theory, that minimal degree algebraic curve, in the unrefined case, is the mirror curve of the geometry under consideration. We thus propose that, in the refined case, our minimal degree spectral curve is the \textit{mirror of our refined geometry}.

\subsection{Limiting shape}

In \cite{eynchain}, the functions $x_s(z)$ and $r_s(z)$ have been introduced as the limit resolvents of matrices $M_s$, namely, if one eliminates $z$ between $x_s(z)$ and $r_s(z)$, we have almost by definition
\be
\left< \tr \frac{1}{x_s(z)-M_s} \right>^{(0)} \sim r_{s+\frac{1}{2} }(z),
\ee
where the upperscript $\langle\,.\,\rangle^{(0)}$ means the planar part of the expectation value, \textit{i.e.}, the leading $\ln t$ term. On the other hand, the resolvent is the Stieljes transform of the density $\rho_s$ of eigenvalues of $M_s$ which means that the density of eigenvalues of $M_s$ is the discontinuity of the resolvent across its cut
\be
\rho_s(x_s(z)) = \frac{1}{2\pi i}\,(r_{s+\frac 1 2}(z_-)-r_{s+\frac 1 2}(z_+))
\ee
where $z_-$ and $z_+$ mean the value of $z$ on the right and on the left of the cut.

\smallskip
By construction, the eigenvalues of $M_s$ are the $h_i(s)$ defined in \eqref{defhis}, and they represent the heights of the slice $s$ of the plane partition, or in other words, $h_i(s)$ is the position of the $i^{\rm th}$ particle at time $s$. Equivalently, it is the position of the $i^{\rm th}$ non--intersecting line at time $s$, hence, the position of the $i^{\rm th}$ vertical edge of the plane partition $\pi$ at time $s$. $\rho_s(h)$ is the density of particles at time $s$, \textit{i.e.}, the density of non--intersecting lines at time $s$, corresponding to the density of vertical edges at time $s$. See figure \ref{limitshape2}. Therefore, from the knowledge of the functions $x_s,\, r_s$, we can recover the limiting shape of our plane partition. This has been done in a general setting in \cite{eynard1}, and corresponds (in the unrefined case) to the limit shape found by \cite{Logan, Cohn, Kenyon, KOS, boutillier}.

\begin{figure}[h]
\begin{center}
\includegraphics[scale=0.7]{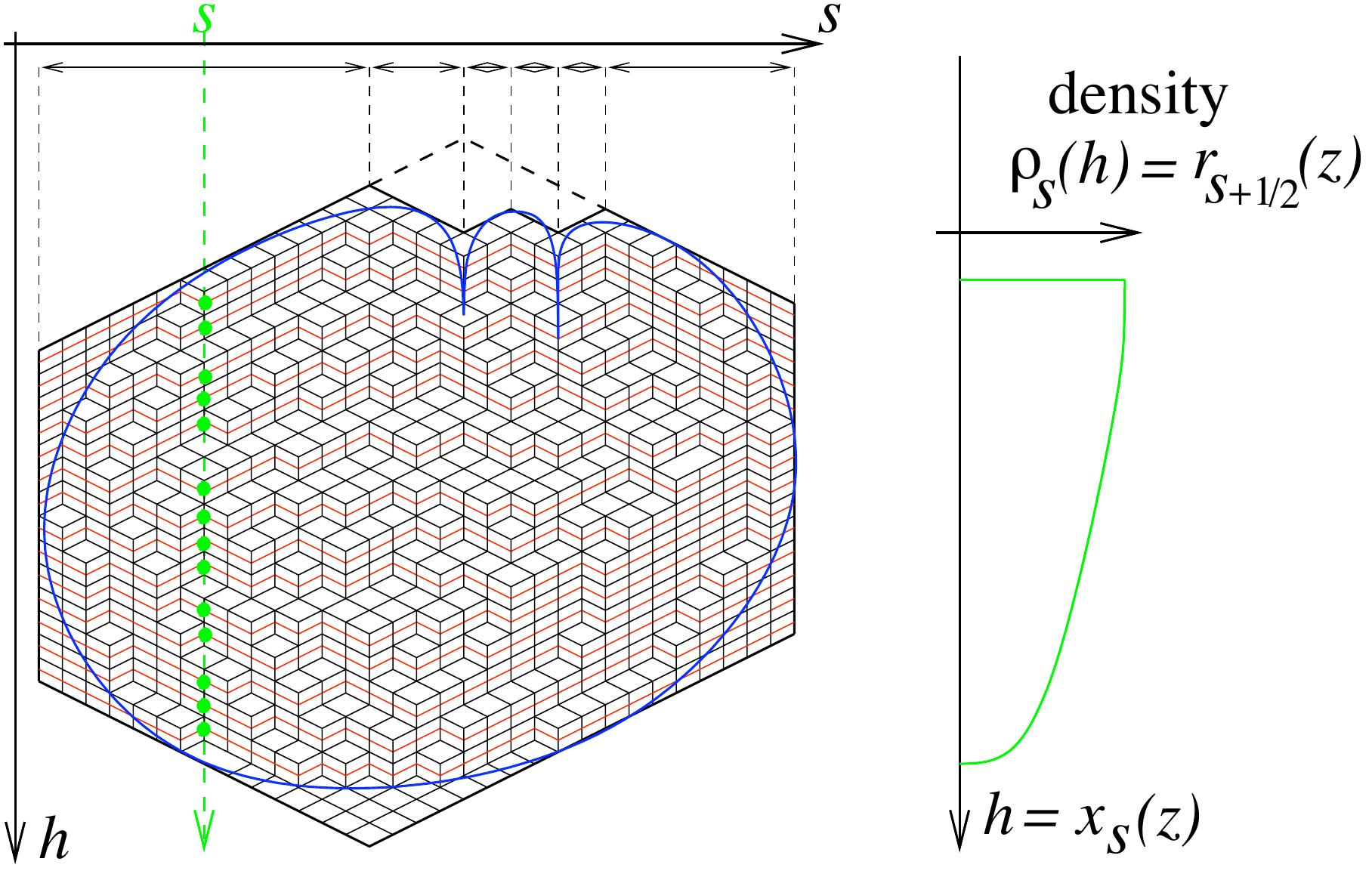}
\end{center}
\caption{From the knowledge of the functions $x_s,\, r_s$ (i.e. the spectral curve) one can recover the density $\rho_s(h)$ of eigenvalues of the matrix $M_s$, and from there the limiting shape of the plane partition can be obtained. The arctic circle is obtained by Legendre transform of the limiting shape.\label{limitshape2}}
\end{figure}

\subsubsection{Reminder unrefined case}

\begin{figure}[h]
\begin{center}
\includegraphics[scale=0.7]{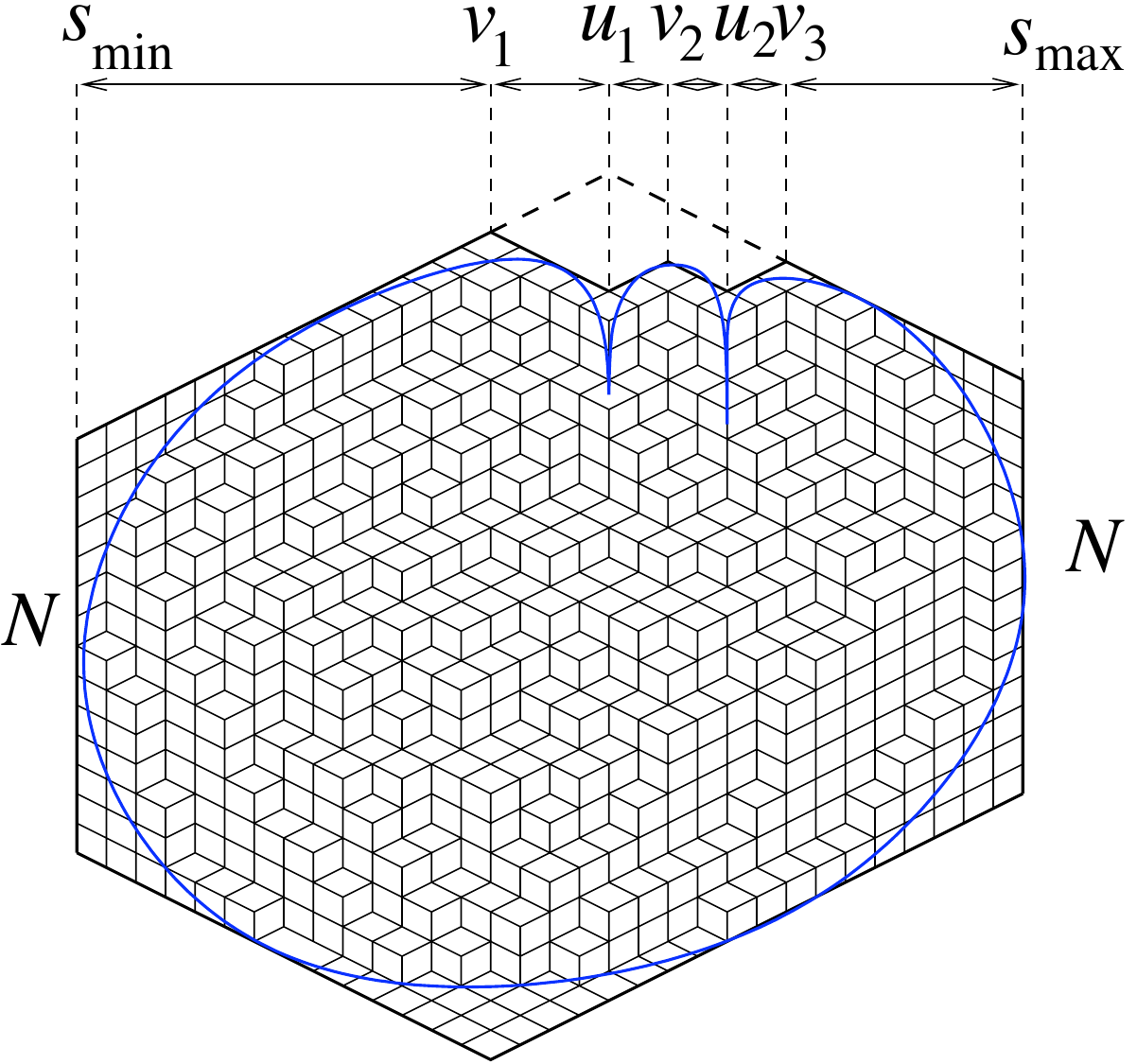}
\end{center}
\caption{The arctic circle is tangent to all sides of the domain, and has cusps at outer corners of $\lambda$.\label{limitshape3}}
\end{figure}

In \cite{Logan, Cohn, Kenyon, KOS, boutillier}, it has been found that the limit shape of the plane partition is such that there is a ``frozen" region (where the density of particles = vertical edges is constant) and a ``liquid" region. The curve separating the frozen region from the liquid region is called the ``arctic circle", see figure \ref{limitshape3} and figure \ref{limitshape4}.

\begin{figure}[h]
\begin{center}
\includegraphics[scale=0.7]{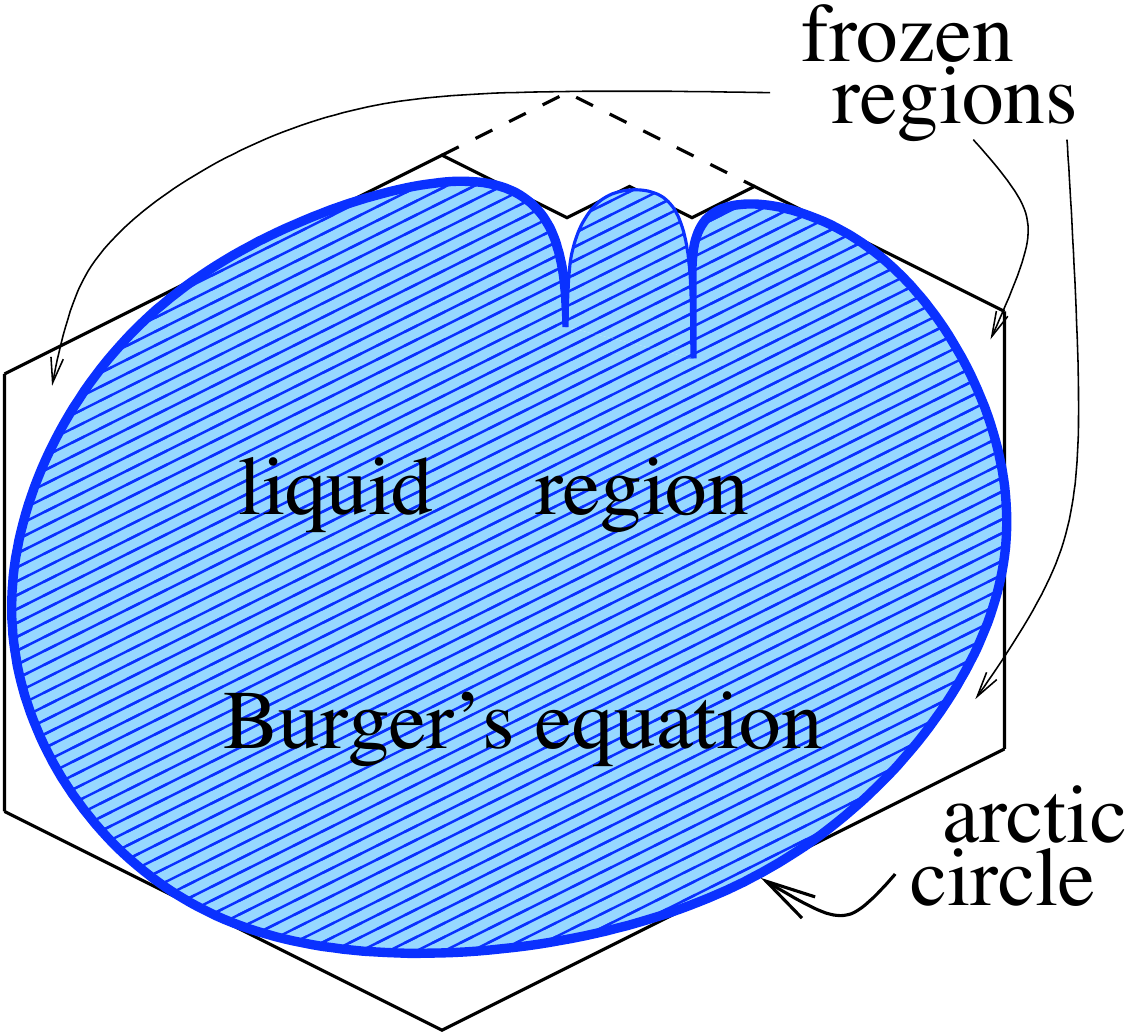}
\end{center}
\caption{The statistical behavior of the surface in the liquid region is governed by the Burger's equation.\label{limitshape4}}
\end{figure}

Furthermore, in \cite{KOS}, it has been shown that the arctic circle is a real algebraic curve, inscribed in the domain of the plane partition (the polygon bounded by the profile $r_\lambda(s)$ on the top, and by the lines $s=s_{\rm min}$ and $s=s_{\rm max}$ on the sides, and by the lines $h=N-s_{\rm min}+s/2$ and $h=N+s_{\rm max}-s/2$ at the bottom), tangent to all sides, and with the minimal possible degree given by
\be
{\rm degree}=\frac{1}{3}\,(5+2\#\{u_i\}+\#\{v_i\})
\ee

For example, if $\lambda=\emptyset$ is the empty partition, the profile $r_\lambda(s)=|s|/2$ has only one inner corner $v_1=0$, and no outer corners, \textit{i.e.}, $\#\{u_i\}=0$ and $\#\{v_i\}=1$. In this case, the arctic circle is an algebraic curve of degree $2$, that is the unique ellipse inscribed in the hexagon. See figure \ref{limitshape3} for the typical shape of the arctic circle. 

\medskip

In general, the arctic circle found in \cite{KOS} is called a Harnack curve and has many nice properties, see \cite{KOS} for details. In addition, it has been shown that, within the liquid region, the  small $\ln t$  statistical behavior of the surface of the plane partition is governed by the Burger's equation. See figure \ref{limitshape4}. We believe that the loop equations of the corresponding matrix model, are equivalent to the Burger's equation, although this has never been carefully proven.

\subsubsection{Some general remarks about the limit shape}

The general properties of the arctic circle curve have been studied for any weight $w(h,s)$ in the domain in \cite{eynard1}.
In particular, it has been shown that  the arctic circle is always tangent to the boundaries of the domain, and has vertical tangents at times $s=u_i$, \textit{i.e.}, at outer corners of the boundary. For any weight $w(h,s)$ the limit shape is always continuous and has a continuous tangent at everypoint except at cusps. In particular for our case, this means that the arctic circle is continuous and has a continuous tangent at inner corners $s=v_i$.
 \medskip

It has been also shown in \cite{eynard1} that the arctic circle curve and the spectral curve $(x_s(z),r_s(z))$ are related by Legendre transform, namely, let $h=\hat x(s)$ be the equation of the arctic circle in the $(h,s)$ plane.
Remember that the spectral curve is
\be
X_s(z) = t^{-x_s(z)} = X_{v_i}(z)\,\frac{t^{\frac{v_i-s}2}+t^{\frac{s-v_i}2}Y_i(z)}{1+Y_i(z)},
\ee
of the general form
\be
X_s(z) = t^{\frac{-s}2}\,\left(A(z) + t^s\,B(z)\right),
\ee
i.e. $t^{s/2} X_s(z)$  is linear in $t^s$.
The value at the extremum given by $\d X_s(z)/\d z=0$ is thus
\be
t^{-\hat x(z)} = t^{\frac{-s}2}\,\left(\mathop{{\rm extremum}}_z\,\,(A(z) + t^s\,B(z))\right) = t^{\frac{v_i-s}2}\,\,{\rm Legendre}(A\circ B^{-1})(t^s),\,\, 
\ee
\textit{i.e.}, it is a Legendre transform in the variable $t^s$.
\medskip

The reverse Laplace transform can also be explicitly performed  as well, let  $t^{-h}= t^{-s/2}\,F(t^s)$ be the equation of the arctic circle in the $(h,s)$ plane. We can recover the spectral curve $(x_s(z),r_s(z))$ by choosing the spectral variable $z$ to be (locally) the argument of $F$, namely
\be
\left\{\begin{array}{l}
X_s(z)=t^{-x_s(z)} = t^{-s/2}\,\,\left(F(z)-t^s\,F'(z)\right) \cr
Y_i(z) = t^{v_i}\,\frac{F'(z)}{F(z)}.
\end{array}\right. 
\ee

Note that equations \eqref{eqzeroYi} and \eqref{eqpoleYi}, are precisely the conditions that the limit shape is tangent to the boundary of the domain.


\section{Topological recursion and mirror symmetry, ``remodeling the refined B-model"}

The chain of matrices has been completely solved in the following sense: 

$\bullet$ We can write loop equations, \textit{i.e.}, identities among correlation functions are obtained by integration by parts. Loop equations for general chain of matrices have been written in \cite{eynchain}.

$\bullet$ If we assume that all correlation functions have an expansion in some small parameter $g_s$, such that the connected part of the $n-$point function behaves as $g_s^{2g-2+n}$, then the general solution of loop equations for the chain of matrices has been derived in \cite{eynpratschain}. It has been found in \cite{eynpratschain} that the solution of loop equations is given by the topological recursion defined in \cite{EO}, applied to the spectral curve we have just described in the previous section.

\medskip
The axiomatic definition of the topological recursion has been introduced in \cite{EO}. It is a recursion starting from the data of a Riemann surface ${\cal C}$ and two analytical functions $x:{\cal C}\to \mathbb CP^1,\,y:{\cal C}\to \mathbb CP^1$. It associates some sequences of symmetric meromorphic differential forms $W_n^{(g)}$ on $\otimes^n T^*({\cal C})=T^*({\cal C})\otimes \dots\otimes T^*({\cal C})$, and some complex numbers $F_g({\cal S})=W_0^{(g)}\in\mathbb C$ to a spectral curve ${\cal S}=\{{\cal C},x(z),y(z)\}$.
The recursion goes on $-\chi=2g+n-2$, \textit{i.e.} $W_n^{(g)}$ is computed in terms of $W_{n'}^{(g')}$ with $2g'+n'-2<2g+n-2$ i.e. $-\chi'<-\chi$. We refer to \cite{EO} for details.

\medskip

The numbers $F_g({\cal S})$ are sometimes called the symplectic invariants of ${\cal S}$. The main result of \cite{eynpratschain} is that the formal small $g_s=\ln t$ expansion of the logarithm of the partition function of a chain of matrices, is given by the symplectic invariants of its spectral curve
\be
\ln Z = \sum_{g=0}^\infty (\ln t)^{2g-2}\,F_g(\{\ln X_s,\ln Y\}),
\ee
where $F_g$ is the $g^{\rm th}$ symplectic invariant of the spectral curve $\{{\cal C},\ln X_s(z),\ln Y(z)\}$
(and it is noticed in \cite{eynpratschain} that symplectic invariance implies that $F_g(\{{\cal C},\ln X_s,\ln Y\})$ is independent of $s$, as expected). In other words, knowing the spectral curve, one can recover the full $\ln t$ power series expansion of the partition function, by applying the topological recursion to the spectral curve.

\bigskip

We shall not enter detailed computation of symplectic invariants of the spectral curve we have just found, but we just mention that this statement is the {\bf refined version of the ``re--modeling the B--Model"} of Mari\~no, Bouchard, Klemm, Pasquetti \cite{Marino:2006hs,Bouchard:2007ys} (at the moment for the refined topological vertex and the refined strip geometry). This seems to imply that mirror symmetry extends to the refined version of Gromov--Witten theory.

\section{Examples}\label{ex}

\subsection{The hexagon}

Consider the case where the 3 partitions bounding our plane partition are all empty partitions
\be
\lambda, \mu,\nu=\emptyset.
\ee
For the moment we consider that $s_{\rm min}=-a<0$, $s_{\rm max}=b>0$ and $N$ are finite.
The profile of $\lambda=\emptyset$ is:
\be
r_\lambda(s)=|s|/2,
\ee
and thus the domain for our non--intersecting paths is the hexagon bounded by the 6 lines:
\be
{\rm hexagon}\, = \,\,\,
\left\{\begin{array}{l}
s=-a \cr
s=b \cr
h=-s/2 \cr
h=s/2 \cr
h=N+a+s/2 \cr
h=N+b-s/2 \cr
\end{array}\right.
\qquad
\includegraphics[scale=0.6]{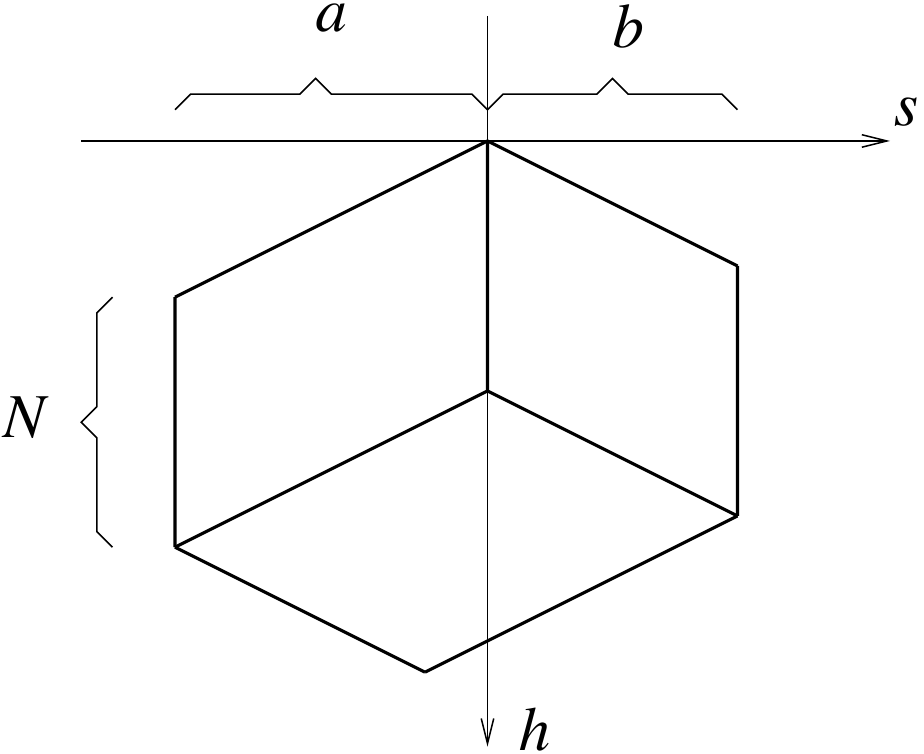}
\ee

The corresponding matrix model of \eqref{eqZi-1i} reduces to
\be\label{Zhex1}
{\cal Z}=
\int_{(i H_{N})}  dR\,\,\,\,
\re^{\mbox{\scriptsize Tr}\,R(M_{b}-M_{-a})}
\,\re^{a\ln t\, \mbox{\scriptsize Tr}\, M_{-a}}
\,\re^{(b-1)\ln q \,\mbox{\scriptsize Tr}\, M_{b}} 
\,\,\frac{\det \left(-t\, e^{-R};t \right)_{\infty}}{\det \left(-t^{a+1}\, e^{-R};t \right)_{\infty}}\frac{\det \left(-e^{R};q\right)_{\infty}}{\det \left(-q^{b}\, e^{R};q\right)_{\infty}}
\ee
where $M_{-a}={\rm diag}(a/2,a/2+1,\dots,a/2+N-1)$ and $M_{b}={\rm diag}(b/2,b/2+1,\dots,b/2+N-1)$, \textit{i.e.},
\be
M_{b}-M_{-a}=\frac{b-a}{2}\,{\rm Id}
\ee
and
\be
\tr\, M_{-a} = \frac{N(N-1)}{2} + \frac{N\,a}{2}
\quad , \quad
\tr\, M_{b} = \frac{N(N-1)}{2} + \frac{N\,b}{2}
\ee
Our matrix model therefore reduces to a 1-matrix integral
\be\label{ZHexagonabN}
{\cal Z}
\propto
\,\int_{(i H_{N})}  dR\,\,\,\,
\re^{\frac{b-a}{2}\,\,\tr\, R}
\,\,\frac{\det \left(-t\, e^{-R};t \right)_{\infty}}{\det \left(-t^{a+1}\, e^{-R};t \right)_{\infty}}\frac{\det \left(-e^{R};q\right)_{\infty}}{\det \left(-q^{b}\, e^{R};q\right)_{\infty}}.
\ee

\medskip

The recipe of section \ref{sectionspcurvealg} for finding the spectral curve, gives:

\be\label{eqhexagoneqs}
\left\{
\begin{array}{l}
t^{-x_s(z)} = X_s(z) = u(z)\,\,\frac{t^{-s/2}\,t^{s/2}\,Y(z)}{1+Y(z)} \qquad ,\,\, -a\leq s\leq 0 \cr
\cr
q^{-x_s(z)} = (X_s(z))^{1/\beta} = u(z)^{1/\beta}\,\,\frac{q^{-s/2}\,q^{s/2}\,Y(z)}{1+Y(z)} \qquad ,\,\, 0\leq s\leq b \cr
\cr
{\rm where}\, u(z)\, {\rm and}\, Y(z)\,\,{\rm must\, satisfy:}
\cr
u=1 \qquad , \,{\rm when}\, Y=\infty \cr
u=1 \qquad , \,{\rm when}\, Y=0 \cr
u=t^{-a-N} \qquad , \,{\rm when}\, Y=0 \cr
u=t^{-b-N} \qquad , \,{\rm when}\, Y=\infty \cr
u=0 \qquad , \,{\rm when}\, Y=-1 \cr
u=\infty \qquad , \,{\rm when}\, Y=-t^a \cr
u=\infty \qquad , \,{\rm when}\, Y=-q^{-b} \cr
\end{array}
\right.
\ee
The minimal assumption is to assume that the variable $z$ lives in $\mathbb CP^1$, and $u(z)$ and $Y(z)$ have only the number of singularities required above: $Y(z)$ has two poles and two zeroes, and $u(z)$ has one pole and one zero of order 1, and one pole and one zero of order $\beta$. Up to a reparametrization of $z$, we may assume that three of the points are $0,1,\infty$, we make the Ansatz:
\be
Y(z) = -\,C\,\frac{(z-\alpha)(z-\gamma)}{(z-\delta)(z-\epsilon)}
\qquad , \quad
u(z) = r\,\,\frac{z}{1-z}\,\,\left(\frac{1}{1-w\,z}\right)^\beta.
\ee
The 7 unknown coefficients $C,\alpha,\gamma,\delta,\epsilon,r,w$ are then determined by the conditions \eqref{eqhexagoneqs}, which read in terms of this ansatz
\be
\begin{array}{l}
u(\delta)=1 \cr
u(\gamma)=1 \cr
u(\alpha)=t^{-N-a} \cr
u(\epsilon)=t^{-N-b} \cr
Y(1)=-t^a \cr
Y(1/w) = -q^{-b} \cr
Y(0)=Y(\infty)=-1
\end{array}
\ee
One can easily check that these 8 equations are not independent, only 7 of them are independent, and thus they allow us to determine the 7 unknowns. Plots of the corresponding arctic circle  for various values of $\beta$ generated by a Mathematica code are displayed in figure \ref{fighexagon}.
\begin{figure}[h]
\begin{center}
\includegraphics[scale=0.2]{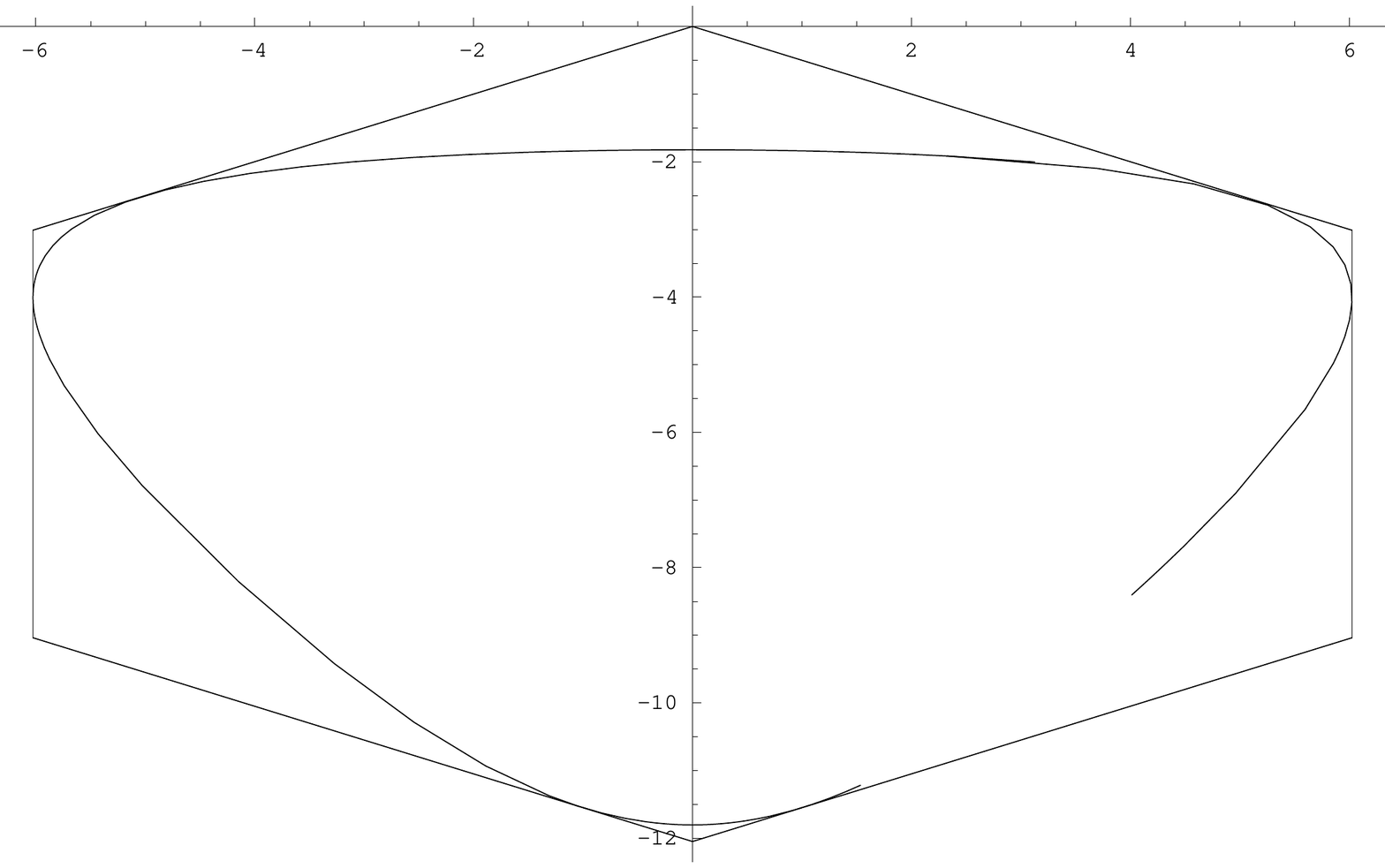}
\includegraphics[scale=0.2]{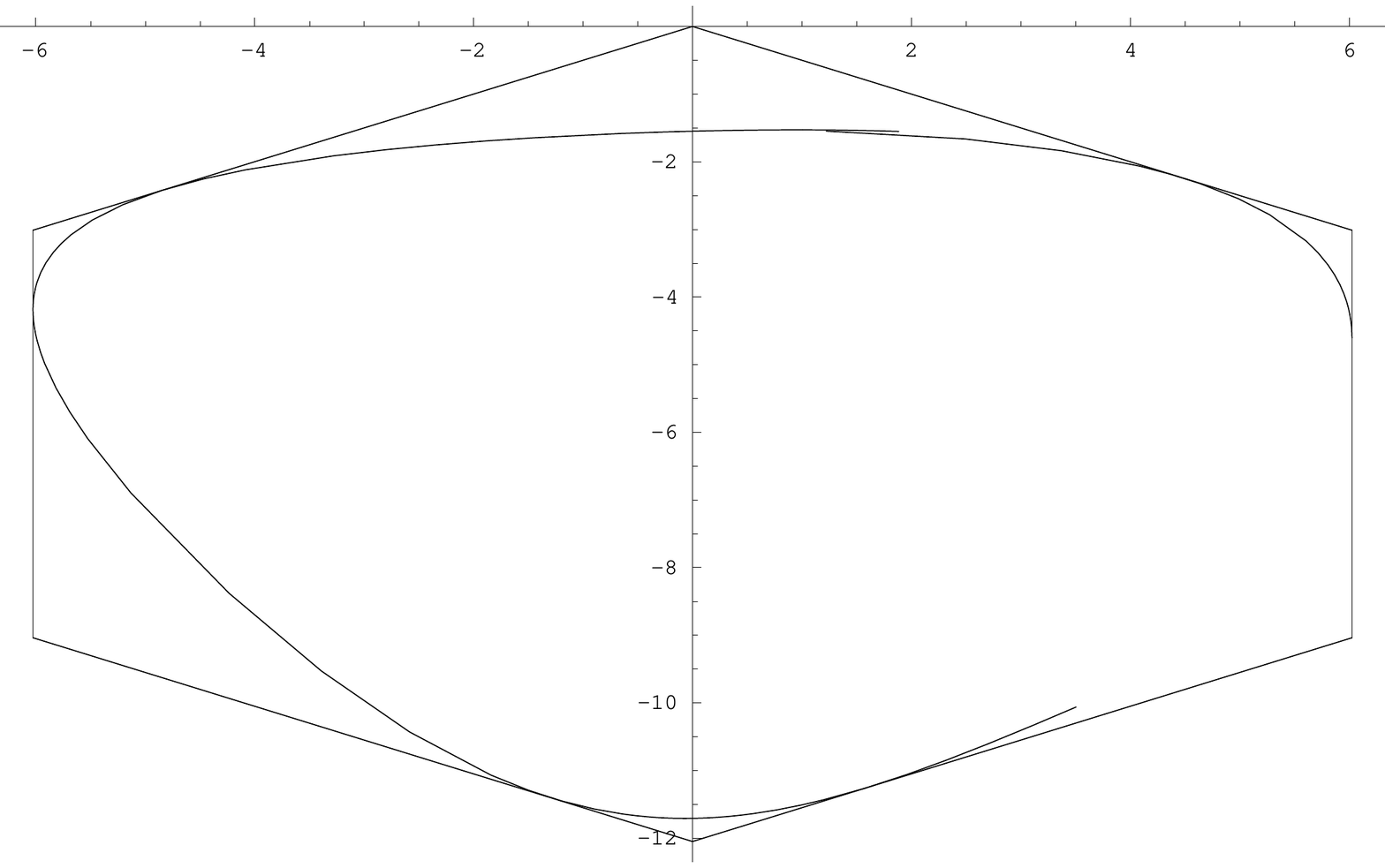}
\includegraphics[scale=0.2]{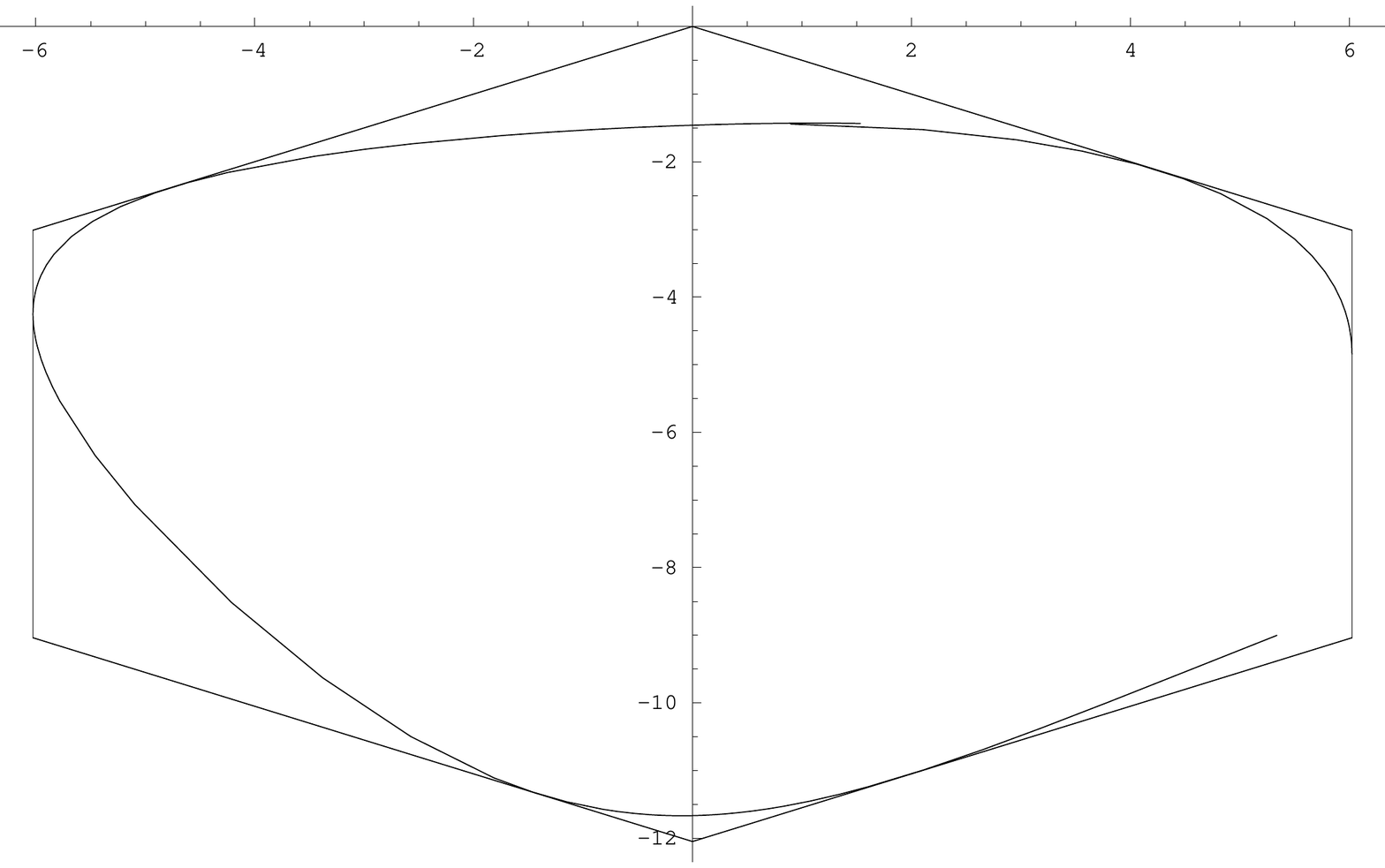}
\includegraphics[scale=0.2]{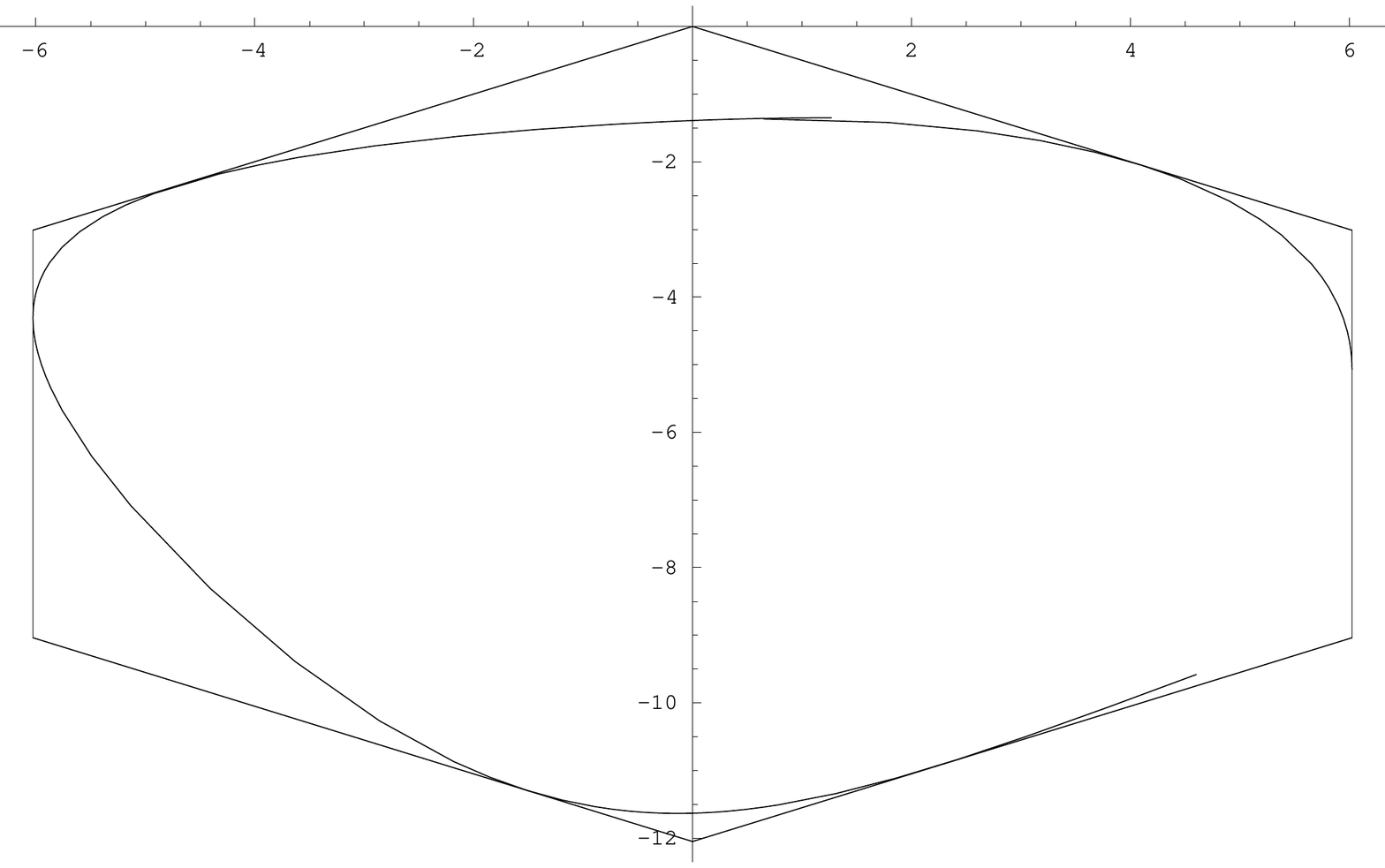}
\includegraphics[scale=0.2]{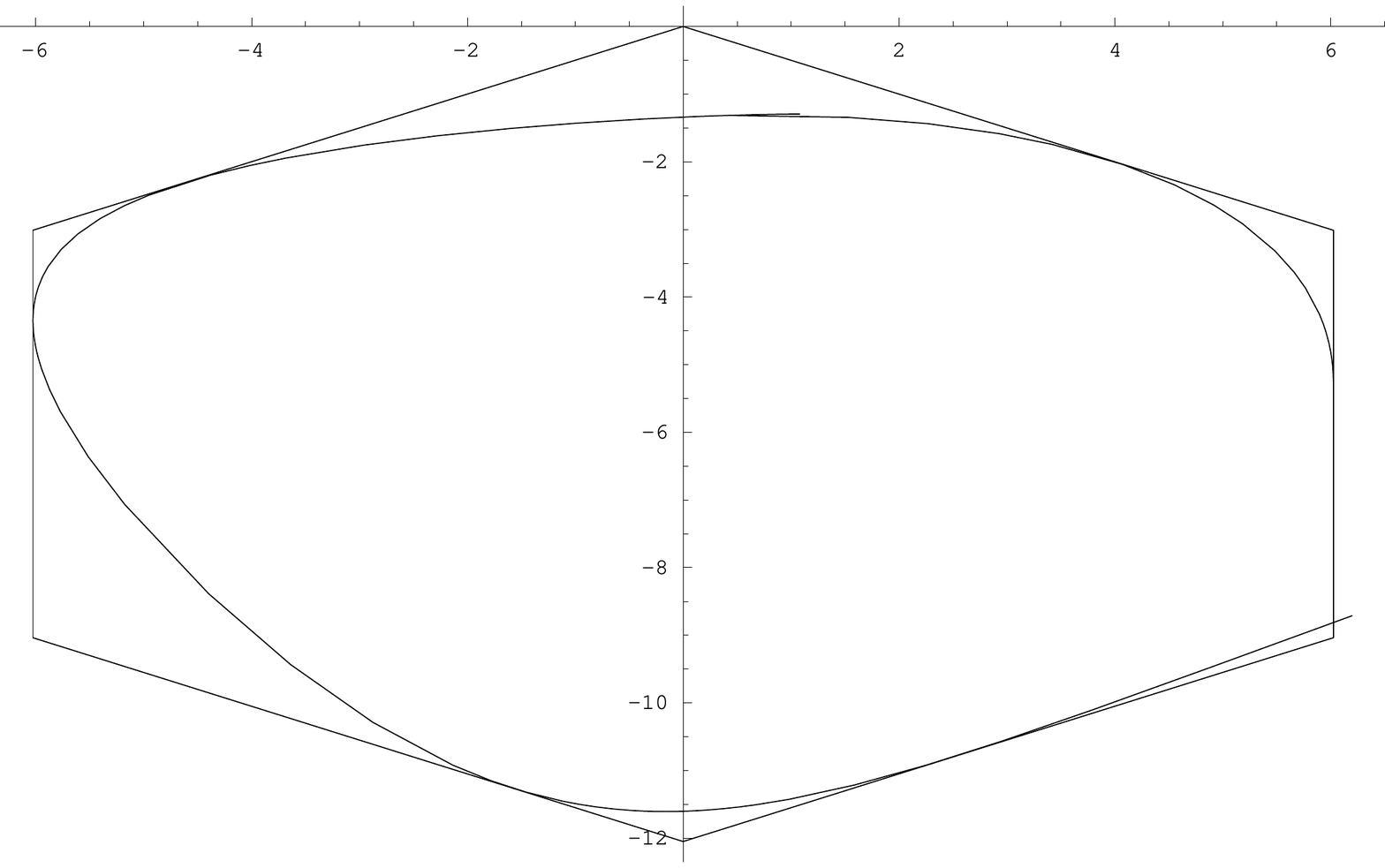}
\includegraphics[scale=0.2]{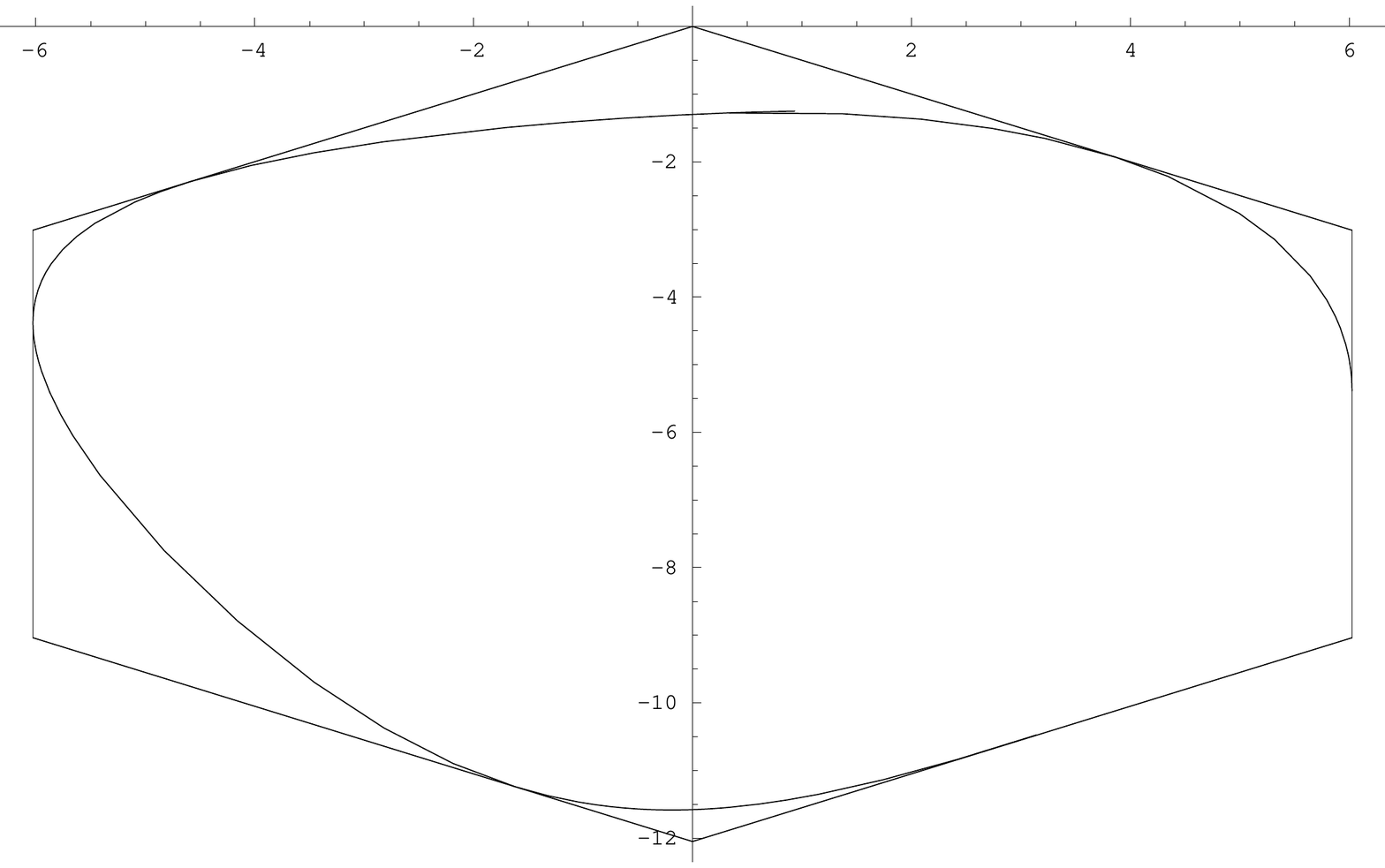}
\end{center}
\caption{Numerical plots of the arctic circle for the hexagon with $a=b=N$, and $\beta=1,2,3,5,10,50$. One side of the arctic circle is missing due to non real valuedness of the parameter in the parametric plot.\label{fighexagon}}
\end{figure}

\subsubsection{Sending the size to $\infty$}

The refined topological vertex is recovered in the infinite length limit of the hexagon. It turns out it is more convenient to start first to keep $a$ and $b$ finite and send $N\to\infty$. In this limit, $Y(z)$ needs to have only one pole and one zero.
Upon a reparametrization of $z$, we may choose
\be
Y(z)=-z
\ee
The recipe for the spectral curve \eqref{eqhexagoneqs} is rewritten as
\be
\begin{array}{l}
u(1)=0 \cr
u(0)=t^{-a} \cr
u(\infty)=t^{-b} \cr
u(t^a)=\infty \cr
u(q^{-b}) = \infty \cr
\end{array}
\ee
that gives the spectral curve
\be
\left\{\begin{array}{l}
Y(z)=-z \cr
u(z) = t^{-b}\,\,\frac{(1-z)^{1+\beta}}{(t^a-z)\,(q^{-b}-z)^\beta}
\end{array}\right.
\ee
and we have
\be
\left\{\begin{array}{l}
X_s(z) = t^{-b-\frac{s}{2}}\,\,\frac{1-t^{s}z}{t^a-z}\,\,\left(\frac{1-z}{q^{-b}-z}\right)^\beta \qquad\quad s\leq 0\cr
X_s(z) = t^{-b-\frac{s}{2}}\,\,\frac{1-z}{t^a-z}\,\,\left(\frac{1-q^s z}{q^{-b}-z}\right)^\beta \qquad\quad s\geq 0
\end{array}\right.
\ee

See examples of plots of the limit shape in figure \ref{PlotHexNinfty} for various values of $a$ and $b$ keeping $\beta$ fixed in the limit $N\to\infty$.
\begin{figure}[h]
\begin{center}
\includegraphics[scale=0.3]{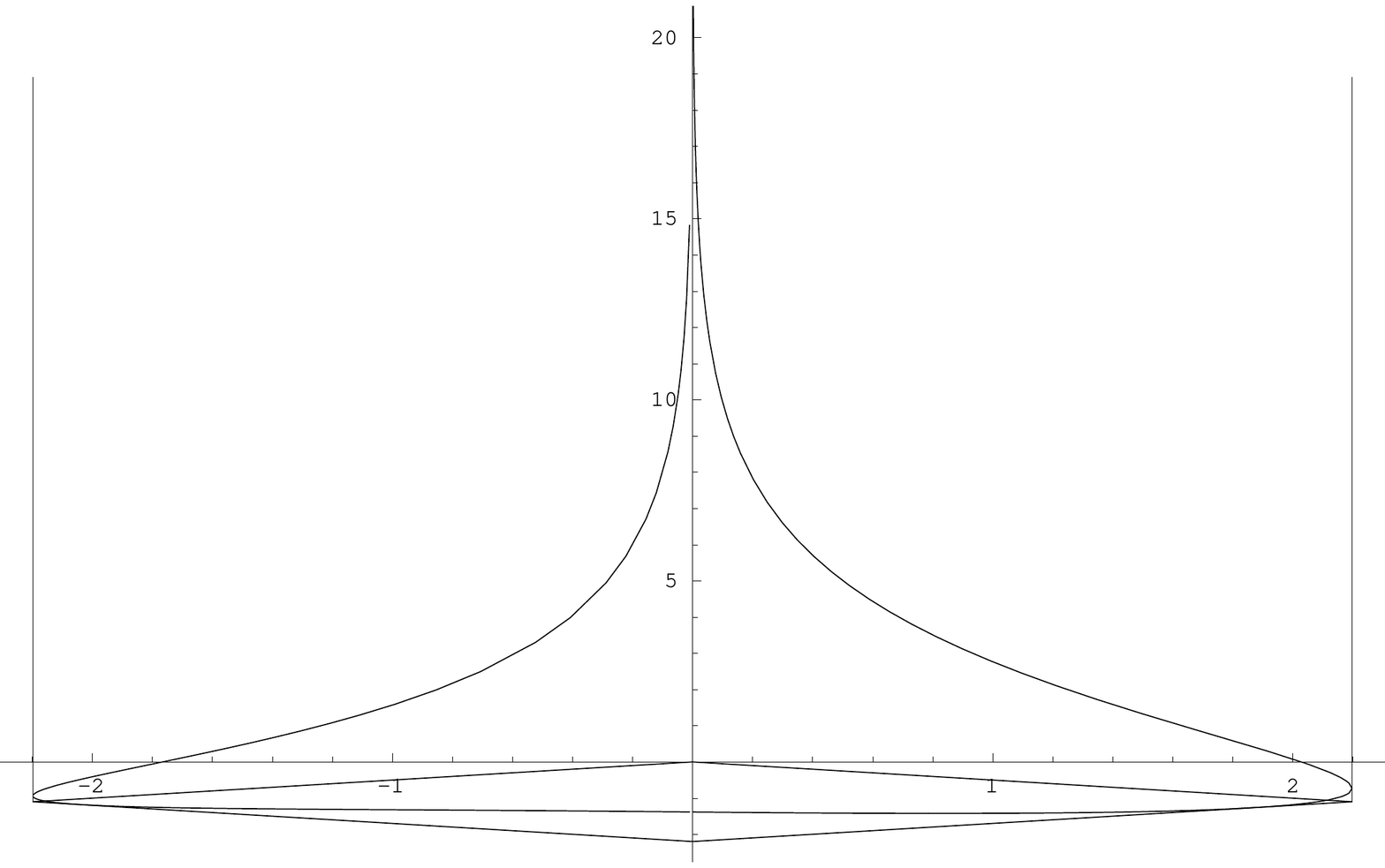}
\includegraphics[scale=0.3]{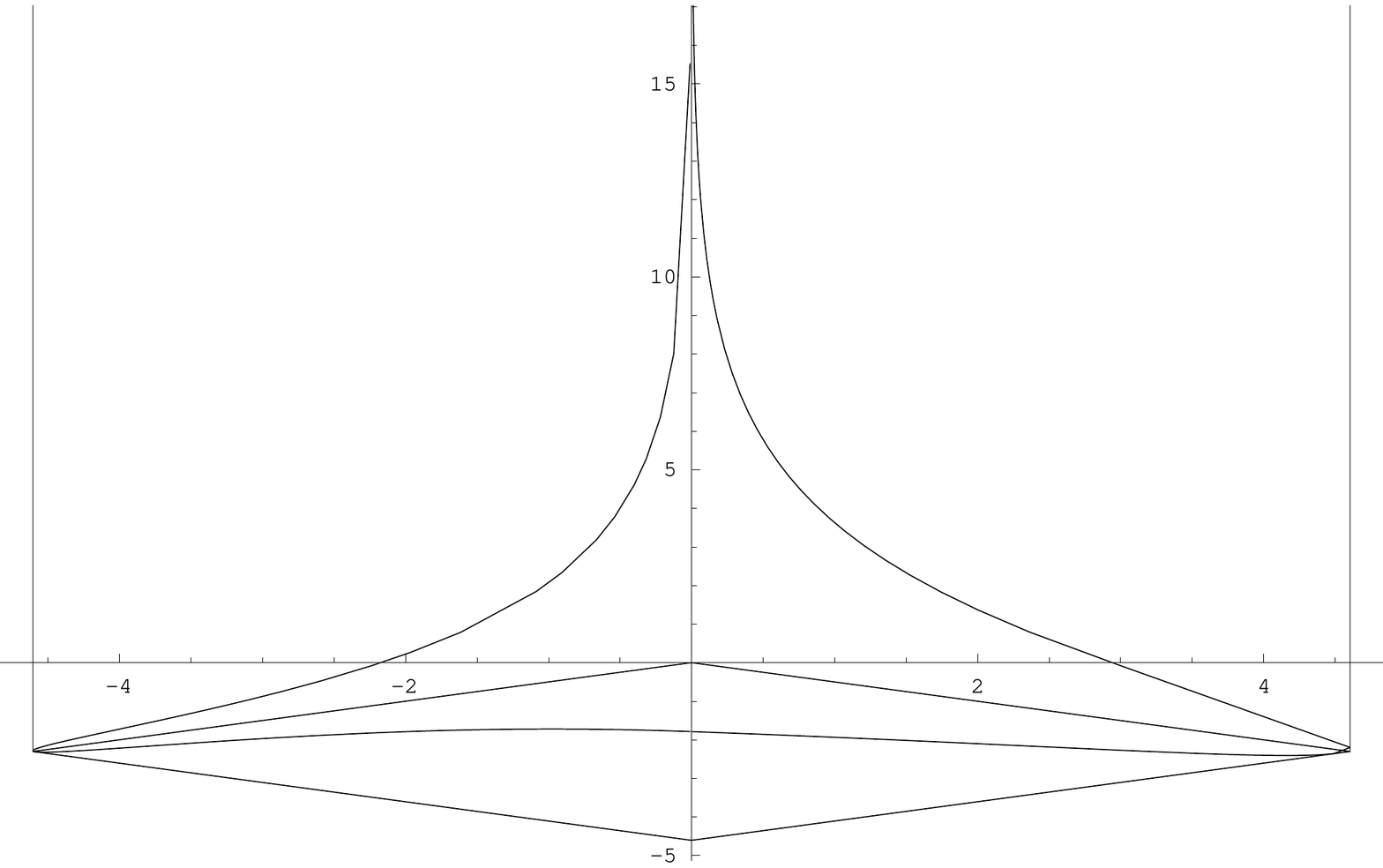}
\includegraphics[scale=0.3]{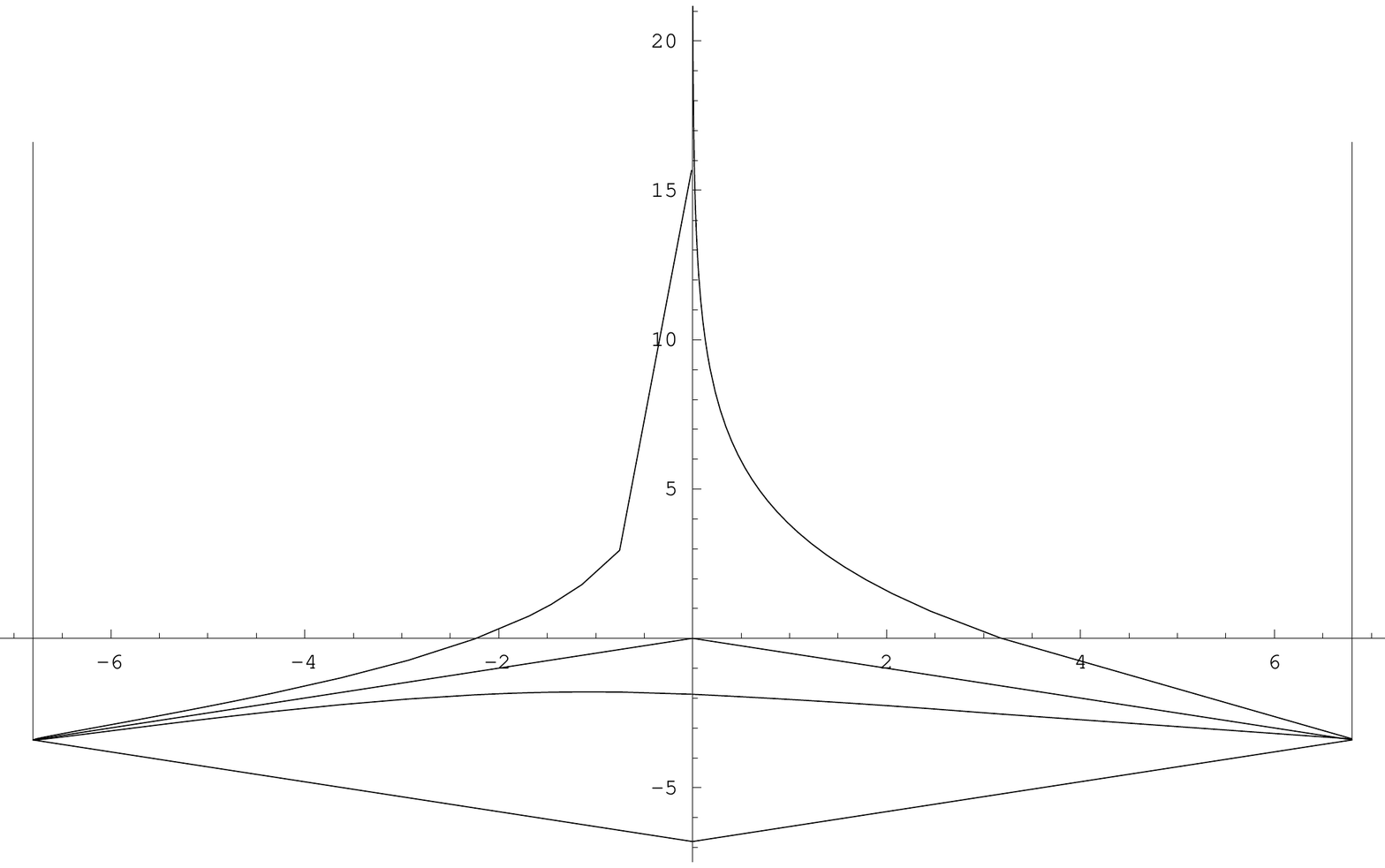}
\includegraphics[scale=0.3]{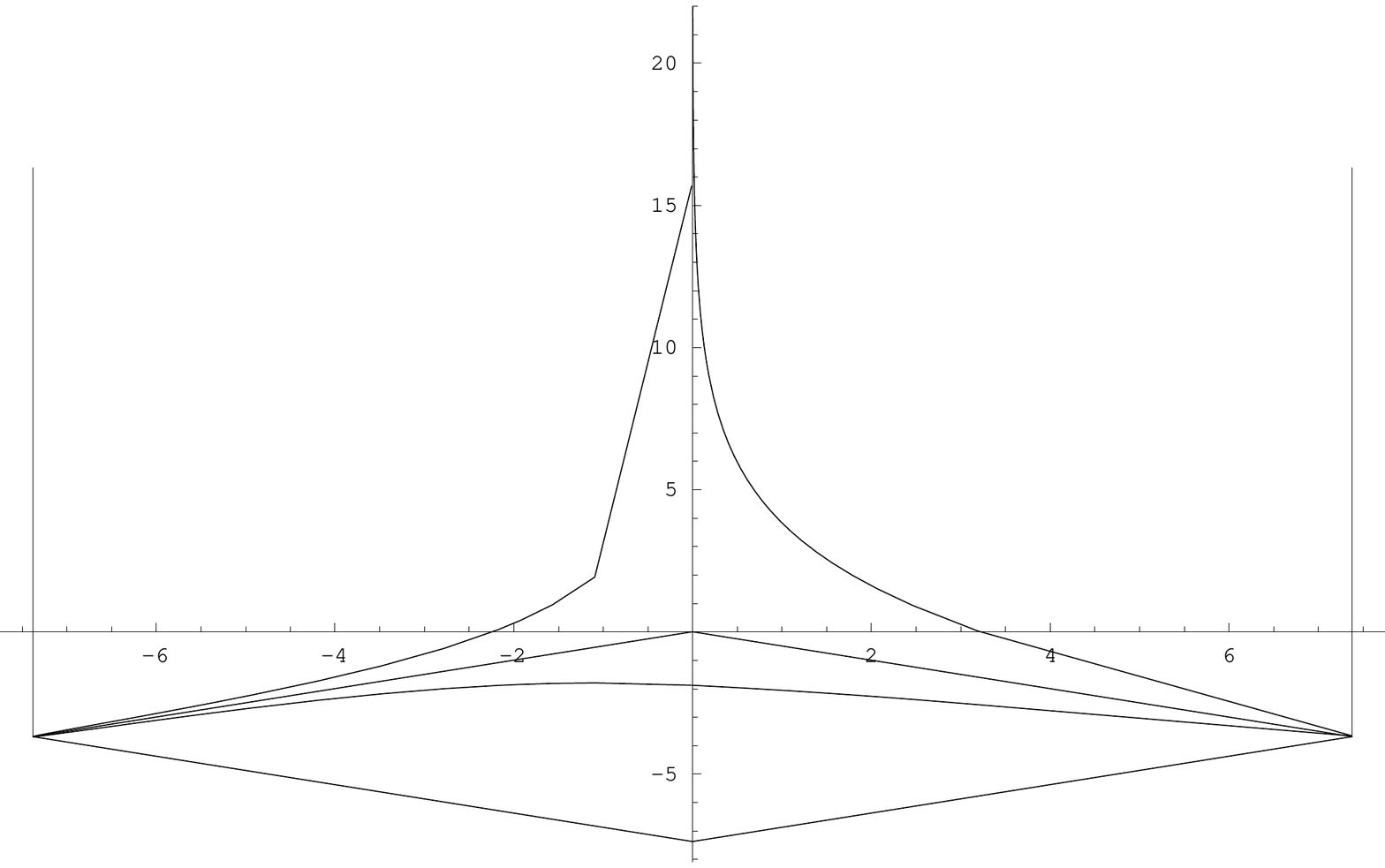}
\end{center}
\caption{with $\beta=2$, we plotted the hexagons $t^a=t^b=1/9$, $t^a=t^b=1/100$, $t^a=t^b=1/900$, $t^a=t^b=1/1600$\label{PlotHexNinfty}}
\end{figure}
Now it is straight forward to send $a\to\infty$ and $b\to\infty$ and we end up with the limit of the spectral curve
\be
\left\{\begin{array}{l}
Y(z)=-z \cr
u(z) = -\,\,\frac{(1-z)^{1+\beta}}{z}
\end{array}\right.
\ee
and we have
\be
\left\{\begin{array}{l}
X_s(z) = -\,t^{-\frac{s}{2}}\,\,\frac{1-t^{s}z}{z}\,\,\left({1-z}\right)^\beta \qquad\quad s\leq 0\cr
X_s(z) = -\,t^{-\frac{s}{2}}\,\,\frac{1-z}{z}\,\,\left({1-q^s z}\right)^\beta \qquad\quad s\geq 0
\end{array}\right.
\ee

This is the spectral curve of the refined topological vertex:
\be\label{spcurverefinedvertex}
{\rm spectral\, curve\, of \, the \, vertex}\quad = \quad
\left\{\begin{array}{l}
y(z)=\ln{Y(z)}=\ln z \cr
x(z) = \ln{u(z)} = \ln{\frac{(1-z)^{1+\beta}}{z}}
\end{array}\right.
\ee

\subsubsection{Other approach to the vertex's matrix model}

Now, let us choose $a=b$, and then send $b\to\infty$, and write the matrix model of \eqref{Zhex1}
\be\label{ZHexagoninfty}
{\cal Z}
\propto
\,\int_{(i H_{N})}  dR\,\,\,\,
\,\,{\det \left(-t\, e^{-R};t \right)_{\infty}}\,\,{\det \left(-e^{R};q\right)_{\infty}}
\ee
which is the refined version of the matrix model obtained by Ooguri--Su\l kowski--Yamazaki \cite{OSY}, also found by Su\l kowski in \cite{Sulk2011}.

\medskip

Let us try to compute directly the spectral curve of the 1-matrix model \eqref{ZHexagoninfty}.
This should be easier, because it is a 1--matrix model, much better studied than the multi matrix model.
However, let us mention that we shall not get the same spectral curve, because our previous spectral curve was defined in terms of resolvents of matrices $M_s$, whereas here, we shall compute the resolvent of $R$. However, both spectral curves should be related by a symplectic transformation, and thus should have the same $F_g$'s, and thus compute the same partition function.

\medskip

The spectral curve of the general 1-matrix model $\int dR\, \re^{-\tr V(R)}$, is the large $N$ limit of the resolvent
\be
W(x) = \ln t\left<\tr \frac{1}{x-R}\right>^{(0)}
\ee
It is well known \cite{DGZ} that it lives on a double sheeted Riemann surface, i.e. 2 copies of the Riemann sphere glued along cuts, and along each cut it must satisfy:
\be
W(x+i0)+W(x-i0)=\ln t\,V'(x),
\ee
Moreover, $W(x)$ must be analytical in the physical sheet, and must behave like
\be
\frac{1}{2\pi i}\,\oint_{{\rm support}} W(x)\,dx =  N
\ee
at large $x$ in the physical sheet.

Here we have:
\be
W(x+i0)+W(x-i0) = \ln t\sum_{k\geq 1} \frac{\re^{-x}\,t^k}{1+\re^{-x}\,t^k}-\ln t\sum_{k\geq 0} \frac{\re^{x}\,q^k}{1+\re^{x}\,q^k}
\ee
That gives in the small $\ln t$ limit:
\be
W(x+i0)+W(x-i0) = \ln{{(1+\re^{x})^\beta\,(1+\re^{-x})}}\, + O(\ln t)
\ee
A solution is clearly
\be
\re^{2W} = \frac{(1+Y)^{1+\beta}}{Y}
\ee
which is the same curve we have found in \eqref{spcurverefinedvertex}.
Notice that in this case $W(x)$ has only logarithmic cuts (it has no algebraic cuts as is customary when potentials are polynomial).

\subsection{The cardioid}

Let us now consider $\lambda$ to be a rectangle, \textit{i.e.}, with 2 inner corners at $v_1$ and $v_2$, and one outer corner $u_1$. We chose the origin of the $s$ axis such that $u_1=v_1+v_2$, and we chose the origin of the $h$ axis so that the upper corner of the bottom rectangle is at $h=0$ (the bottom of the room is at a height $h=0$).
The domain is the following
$$
\includegraphics[scale=0.6]{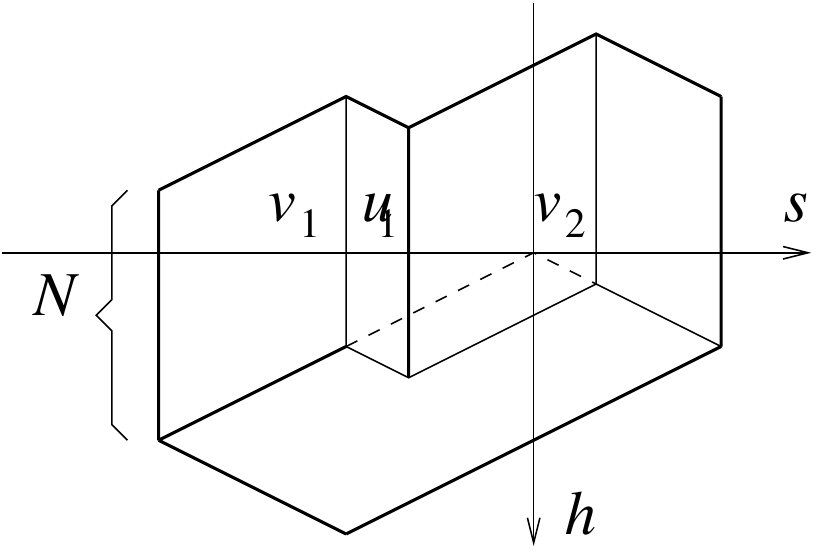}
\qquad
\includegraphics[scale=0.6]{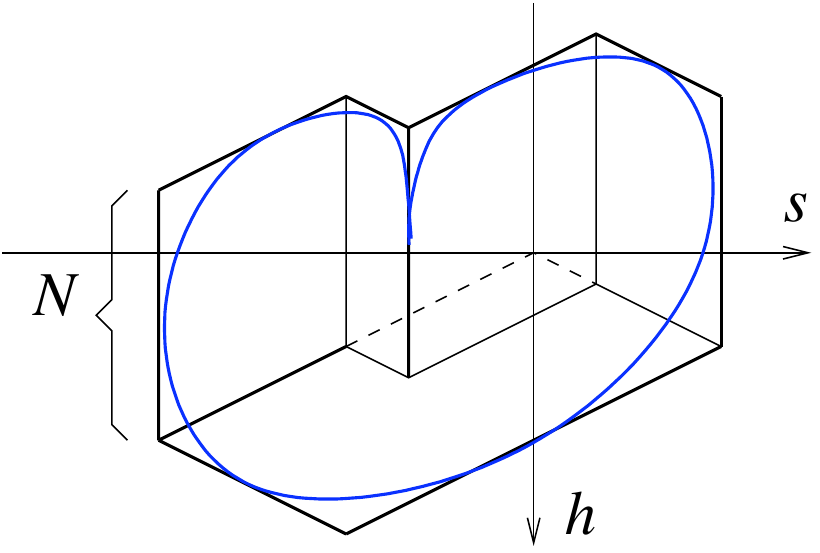}
$$
and thus the limit shape should have 3 tangents of slope $+1/2$, 3 tangents of slope $-1/2$ and 4 vertical tangents. 
All the equations of section \ref{sectionspcurvealg} can be easily written, and the spectral curve can be found at least numerically.
In the unrefined case $\beta=1$, the arctic circle is a degree $3$ algebraic curve, and when $v_1=-v_2=N/2$, this is the famous regular cardioid often shown as an illustration of Okounkov's works \cite{KOS}.

\medskip
However, let us further simplify by  also chosing $N=\infty$. Then, only the tangency conditions at the bottom of the domain remain, 
$$
\includegraphics[scale=0.6]{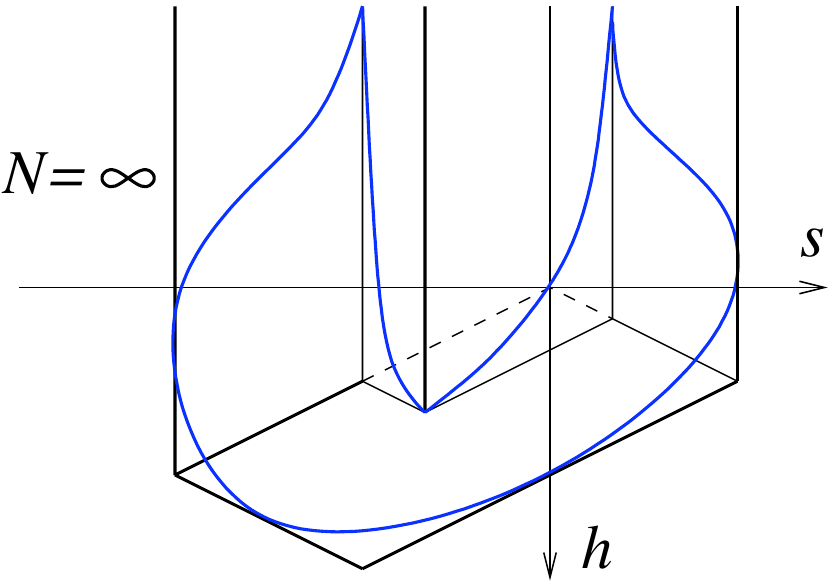}
$$
\textit{i.e.},  we need only one pole and one zero for $Y_1$. We choose
\be
Y_1(z) = -q^{v_1-u_1}\,z
\qquad , \quad 
Y_2(z) = -t^{v_2-u_1}\,z.
\ee
Then we have
\be
\begin{array}{ll}
X_{s}(z) = t^{\frac{v_1-s}2}\,\,X_{v_1}(z)\,\,\frac{1-t^{s-v_1}\,q^{v_1-u_1}\,z}{1-q^{v_1-u_1}\,z} & \qquad u_0\leq s\leq v_1 \cr
X_{s}(z) = t^{\frac{v_1-s}2}\,\,X_{v_1}(z)\,\,\left(\frac{1-q^{s-u_1}\,z}{1-q^{v_1-u_1}\,z}\right)^\beta & \qquad v_1\leq s\leq u_1 \cr
X_{s}(z) = t^{\frac{v_2-s}2}\,\,X_{v_2}(z)\,\,\frac{1-t^{s-u_1}\,z}{1-t^{v_2-u_1}\,z} & \qquad u_1\leq s\leq v_2 \cr
X_{s}(z) = t^{\frac{v_2-s}2}\,\,X_{v_2}(z)\,\,\left(\frac{1-q^{s-v_2}\,t^{v_2-u_1}\,z}{1-t^{v_2-u_1}\,z}\right)^\beta & \qquad v_2\leq s\leq u_2 \cr
\end{array}
\ee
Matching the 2 expressions at $s=u_1$ gives
\be
 X_{v_1}(z)\,\,\left(\frac{1-z}{1-q^{v_1-u_1}\,z}\right)^\beta 
= t^{\frac{v_2-v_1}2}\,\,X_{v_2}(z)\,\,\frac{1-z}{1-t^{v_2-u_1}\,z} 
\ee
Then we want
\be
\begin{array}{l}
X_{u_0}(0)=t^{u_0/2} \cr
X_{u_2}(\infty)=t^{-u_2/2}  \cr
\cr
X_{v_1}(q^{u_1-v_1})^{1/(1+\beta)} = 0 \cr
X_{v_1}(t^{v_1-u_0}q^{u_1-v_1})=\infty \cr
X_{v_1}(1)^{1/\beta}=\infty \cr
\cr
X_{v_2}(t^{u_1-v_2})^{1/(1+\beta)} = 0 \cr
X_{v_2}(t^{u_1-v_2}q^{v_2-u_2})^{1/\beta}=\infty \cr
X_{v_2}(1)=\infty \cr
\end{array}
\ee
We thus find
\be
\begin{array}{lcl}
X_{u_0}(z) &=&  \displaystyle  t^{-\,\frac{u_0}2}\,\frac{(1-q^{v_1-u_1}z)^{\beta}\,}{(1-z)^\beta\,}\,\,\,\frac{(1-t^{v_2-u_1}z)^\beta}{(1-q^{u_2-v_2}t^{v_2-u_1}z)^\beta}\,\,  \cr
&&\cr
X_{v_1}(z) &=&  \displaystyle    t^{-\,\frac{v_1}2}\,\frac{(1-q^{v_1-u_1}z)^{1+\beta}\,}{(1-t^{u_0-v_1}q^{v_1-u_1}z)\,(1-z)^\beta\,}\,\,\,\frac{(1-t^{v_2-u_1}z)^\beta}{(1-q^{u_2-v_2}t^{v_2-u_1}z)^\beta}\,\, \cr
&&\cr
X_{u_1}(z) &=&  \displaystyle   t^{-\,\frac{u_1}2}\,\frac{(1-q^{v_1-u_1}z)\,}{(1-t^{u_0-v_1}q^{v_1-u_1}z)}\,\,\,\frac{(1-t^{v_2-u_1}z)^\beta}{(1-q^{u_2-v_2}t^{v_2-u_1}z)^\beta}\,\, \cr
&&\cr
X_{v_2}(z) &=&  \displaystyle   \,t^{-\,\frac{v_2}2}\,\frac{(1-t^{v_2-u_1}z)^{1+\beta}\,}{(1-q^{u_2-v_2}t^{v_2-u_1}z)^\beta\,(1-z)\,}\,\,\,\frac{(1-q^{v_1-u_1}z)}{(1-t^{u_0-v_1}q^{v_1-u_1}z)}\, \cr
&&\cr
X_{u_2}(z) &=&  \displaystyle   \,t^{-\,\frac{u_2}2}\,\frac{(1-t^{v_2-u_1}z)\,}{\,(1-z)\,}\,\,\,\frac{(1-q^{v_1-u_1}z)}{(1-t^{u_0-v_1}q^{v_1-u_1}z)}\,
\end{array}\label{eqcardioidNinfty}
\ee
See a numerical plot of this spectral curve for $\beta=2$ in figure \ref{figPlotrefinedcardioidbeta2}.
\begin{figure}[h]
\begin{center}
\includegraphics[scale=0.4]{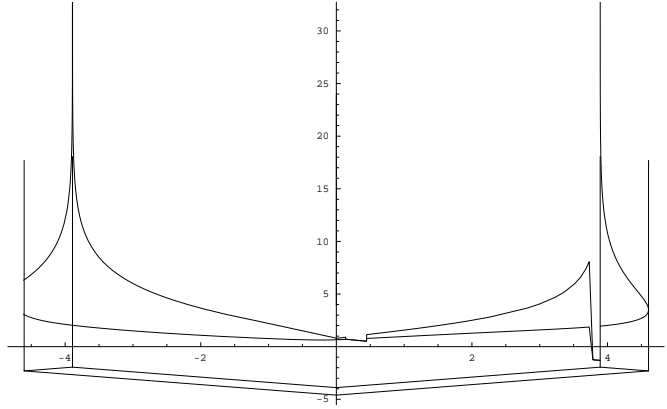}
\end{center}
\caption{Plot of the limit shape obtained from the spectral curve \eqref{eqcardioidNinfty} and with $\beta=2$ (and up to small artifacts from the numerics).  \label{figPlotrefinedcardioidbeta2}}
\end{figure}

\medskip
If then we send $u_0\to -\infty$ and $u_2\to \infty$, we find the spectral curve:
\be
X = \frac{(1+q^{-v_2}Y)^{1+\beta}\,\,(1+t^{-v_1}Y)^\beta}{Y\,\,(1+Y)^\beta}
\ee
This spectral curve is the should--be mirror of  the geometry
$$
\includegraphics[scale=0.7]{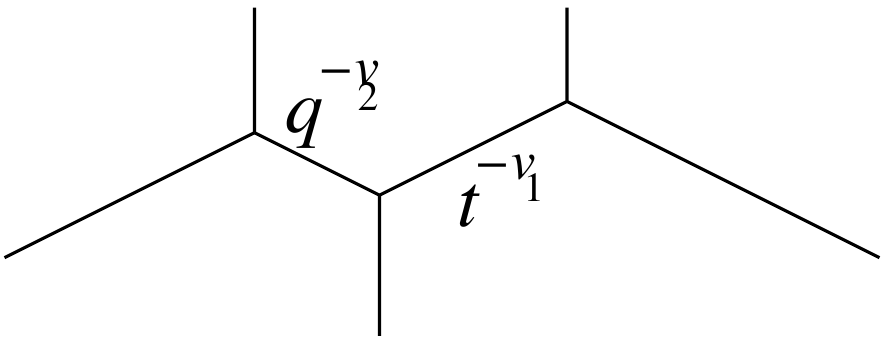}
$$

\subsection{Spectral curve for the general strip}

The above computation for the cardioid with one $u_i$ and two $v_i$'s, can easily be extended to an arbitrary number of $u_i$'s and $v_i$'s.

Again we work at $N\to\infty$, so that $Y$ needs to have only 1 pole and 1 zero, and we choose:
\be
Y_{1}(z) = -z
\ee
and we recall that
\be
Y_{i+1}(z) = t^{v_{i+1}-u_i}\,q^{u_i-v_i}\,Y_i(z),
\ee
which we write
\be
Y_i(z) = - c_i\,z \qquad \quad {\rm with}\,\,\,\,c_{i+1} = t^{v_{i+1}-u_i}\,q^{u_i-v_i}\,c_i.
\ee
similarly to the cardioid case, the minimal degree solution of all constraints of section \ref{sectionspcurvealg} is:
\be
X_{v_i}(z) = t^{-\,\frac{v_i}2}\,\frac{\prod_{j\leq i} (1-c_j\,z)\,\, \,\,\prod_{j\geq i} (1-c_j\,z)^\beta}{\prod_{j\leq i}\,(1-t^{u_{j-1}-v_j}\,c_j\,z)\,\,\prod_{j\geq i}\,(1-q^{u_j-v_j}\,c_j\,z)^\beta} 
\ee
\textit{i.e.}, the mirror of the refined strip is:
\be
X = t^{-\,\frac{v_i}2}\,\frac{\prod_{j\leq i} (1+c_j\,Y)\,\, \,\,\prod_{j\geq i} (1+c_j\,Y)^\beta}{\prod_{j\leq i}\,(1+t^{u_{j-1}-v_j}\,c_j\,Y)\,\,\prod_{j\geq i}\,(1+q^{u_j-v_j}\,c_j\,Y)^\beta} 
\qquad  {\rm with}\,\,\,\,c_{i+1} = t^{v_{i+1}-u_i}\,q^{u_i-v_i}\,c_i.
\ee


\section{Outlook \& Conclusions}

\par{We construct the matrix model of the refined topological vertex and the refined strip geometry that is the building block of the superconformal $SU(N)$ theory with $N_{f}=2N$ matter multiplets. The corresponding spectral curves are found, thus the mirror of the corresponding geometries. Our matrix model differs from the previously proposed ones in the powers of the Vandermonde determinant, known as the $\beta-$ensemble \cite{Dijkgraaf:2009pc}. Unlike the previously constructed matrix models ours is an Hermitian one. The price to pay seems to be that we have a chain matrix model, and $\beta$ lies in the potentials instead of the Vandermonde.     }

\par{Having a matrix model, we first find its spectral curve, and we propose that this spectral curve should be the ``mirror" of the corresponding refined geometry. Then, once the spectral curve is known, all string amplitudes are obtained by the topological recursion, in other words, the refined topological string also satisfies the ``remodeling the B--model" construction.

We find that the ``mirror curve", \textit{i.e.}, the spectral curve of the matrix model, is a certain Legendre transform of the limit shape of the crystal model, in accordance with the unrefined case. For $\beta\in \mathbb Q$ rational, the mirror curve is an algebraic curve.}

\par{The strip geometry we consider in section \ref{toric}. As mentioned two strips can be glued to give rise to the geometry that engineer superconformal gauge theories in 4$d$. In \cite{Kozcaz:2010af}, using this fact and Gaiotto's construction of gauge theories combined with the AGTW relations the refined closed amplitude is connected to chiral $A_{N-1}$ Toda three point function with one of the three primary fields of a special type \cite{Wyllard:2009hg}. Our approach, hence, give a matrix model interpretation of the three point function.  }

\section*{Acknowledgements}

We would like to thank A. Alexandrov, C. Boutillier, A. Brini, M. Mari\~no, P. Su\l kowski  for useful discussions
and correspondence. 
The work of B. E.  is partly supported by the ANR project GranMa ``Grandes Matrices Al\'{e}atoires" ANR-08-BLAN-0311-01, by the European Science Foundation through the Misgam program, by the Quebec government with the FQRNT, and the CERN.


\appendix
\section{Appendix: Useful identities}
In this section we want to collect some of the identities and expressions we have made use of in the derivation of the crystal model for the refined strip geometry. When we have written the generating function for the plane partition we have encountered the following inner products which are written as shown in terms of Schur functions

\begin{align}
&\langle \eta|x^{L_{0}}\prod_{i=1}^{a}\Gamma_{-}(1)t^{L_{0}}|\sigma\rangle=x^{|\eta|}t^{a|\sigma|}s_{\eta/\sigma}(1,t,\mathellipsis,t^{a-1}),\\
&\langle \mu|x^{L_{0}}\Gamma_{+}(1)\prod_{i=1}^{a}q^{L_{0}}\Gamma_{+}(1)|\eta\rangle=x^{|\mu|}q^{a|\mu|}s_{\eta/\mu}(1,q,\mathellipsis,q^{a}),\\
&\langle \nu|x^{L_{0}}\Gamma_{-}(1)\prod_{i=1}^{a}t^{L_{0}}\Gamma_{-}(1)|\mu\rangle=x^{|\nu|}t^{a|\mu|}s_{\nu/\mu}(1,t,\mathellipsis,t^{a}),\\
&\langle\gamma|\prod_{i=1}^{a}q^{L_{0}}\Gamma_{+}(1)|\nu\rangle=q^{a|\gamma|}s_{\nu/\gamma}(1,q,\mathellipsis,q^{a-1}).
\end{align}
We have used the following analytical continuation to express the Schur function with finite number of argument as Schur functions appearing in the topological vertex computations
\be
s_{\lambda}(1,t,\mathellipsis;Q t^{-1},Q t^{-2},\mathellipsis)=s_{\lambda}(1,\mathellipsis,t^{n}),\qquad \mbox{with the choice}\,Q=t^{n+1}.
\ee 
The Schur function on the left hand side of the above expression can be rewritten using 

\be
s_{\lambda}(x,y)=\sum_{\eta}s_{\lambda/\eta}(x)s_{\eta}(y),
\ee
where $x$ and $y$ stand for infinite many arguments. The following well-known identity has been used in the manipulations to show the equivalence 

\be
s_{\lambda/\eta}(q^{-\rho})=(-1)^{|\lambda|+|\eta|}s_{\lambda^{t}/\eta^{t}}(q^{\rho}),
\ee
where $\rho_{i}=-i+\frac{1}{2}$.

\end{document}